\documentclass[aps,prx,twocolumn,superscriptaddress,floatfix,longbibliography]{revtex4-1}

\usepackage{physics}
\usepackage{graphicx}
\usepackage{booktabs}
\usepackage{bbm}
\usepackage{bm}

\usepackage{hyperref}
\hypersetup{
    colorlinks=true,
    linkcolor=blue,     
    urlcolor=blue,
    citecolor=blue
}
\usepackage{soul}
\setcitestyle{open={[},close={]}}

\newcommand{\bsco}{(Bi,Pb)$_2$(Sr,La)$_2$CuO$_{6+\delta}$}
\newcounter{para}

\newcommand{\Qdw}{Q_{\mathrm{DW}}}
\newcommand{\Qdwav}{\bar{Q}_{\mathrm{DW}}}
\newcommand{\Qhs}{Q_{\mathrm{HS}}}

\newcommand{\Qan}{Q_{\mathrm{AN}}}
\newcommand{\Qafzb}{Q_{\mathrm{AFZB}}}
\newcommand{\myvec}[1]{{\bf #1}}
\newcommand{\gap}{\Delta}
\newcommand{\gapc}{\Delta^*}

\begin{document}
\title{Density Wave Probes Cuprate Quantum Phase Transition}

\author{Tatiana A. Webb}
\affiliation{Department of Physics, Harvard University, Cambridge, MA 02138, USA}

\author{Michael C. Boyer}
\affiliation{Department of Physics, Clark University, Worcester, MA 01610, USA}
\affiliation{Department of Physics, Massachusetts Institute of Technology, Cambridge, MA 02139, USA}

\author{Yi Yin}
\altaffiliation[Present address: ]{Physics Department, Zhejiang University, Hangzhou, 310027, China}
\affiliation{Department of Physics, Harvard University, Cambridge, MA 02138, USA}

\author{Debanjan Chowdhury}
\affiliation{Department of Physics, Massachusetts Institute of Technology, Cambridge, MA 02139, USA}

\author{Yang He}
\affiliation{Department of Physics, Harvard University, Cambridge, MA 02138, USA}

\author{Takeshi Kondo}
\altaffiliation[Present address: ]{ISSP, University of Tokyo, Kashiwa, Chiba 277-8581, Japan}
\affiliation{Department of Crystalline Materials Science, Nagoya University, Nagoya 464-8603, Japan}

\author{ T. Takeuchi}
\altaffiliation[Present address: ]{Toyota Technological Institute, Nagoya 468-8511, Japan}
\affiliation{Department of Crystalline Materials Science, Nagoya University, Nagoya 464-8603, Japan}

\author{H. Ikuta}
\altaffiliation[Present address: ]{Department of Materials Physics, Nagoya University, Nagoya 464-8603, Japan
}
\affiliation{Department of Crystalline Materials Science, Nagoya University, Nagoya 464-8603, Japan}

\author{Eric W. Hudson}
\affiliation{Department of Physics, Pennsylvania State University, University Park, PA 16802-6300, USA}

\author{Jennifer E. Hoffman}
\email{jhoffman@physics.harvard.edu}
\affiliation{Department of Physics, Harvard University, Cambridge, MA 02138, USA}

\author{Mohammad H. Hamidian}
\email{m.hamidian@gmail.com}
\affiliation{Department of Physics, Harvard University, Cambridge, MA 02138, USA}

\date{\today}

\begin{abstract}
In cuprates, the strong correlations in proximity to the antiferromagnetic Mott insulating state give rise to an array of unconventional phenomena beyond high temperature superconductivity. Developing a complete description of the ground state evolution is crucial to decoding the complex phase diagram.  Here we use the structure of broken translational symmetry, namely $d$-form factor charge modulations in \bsco{}, as a probe of the ground state reorganization that occurs at the transition from truncated Fermi arcs to a large Fermi surface.  We use real space imaging of nanoscale electronic inhomogeneity as a tool to access a range of dopings within each sample, and we definitively validate the spectral gap $\gap$ as a proxy for local hole doping. From the $\gap$-dependence of the charge modulation wavevector, we discover a commensurate to incommensurate transition that is coincident with the Fermi surface transition from arcs to large hole pocket, demonstrating the qualitatively distinct nature of the electronic correlations governing the two sides of this quantum phase transition. Furthermore, the doping dependence of the incommensurate wavevector on the overdoped side is at odds with a simple Fermi surface driven instability. 
\end{abstract}

\maketitle

\section{Introduction}

In cuprates, high-temperature superconductivity lies between an undoped antiferromagnetic (AFM) insulator and a metal at high hole doping ($p$).  In proximity to the AFM insulator, the strong electronic correlations give rise to a complex phenomenology, including a large spectral gap $\gap$ that opens above $T_c$, and a $k$-space structure lacking a conventional Fermi surface (FS) but described by open arcs \cite{LeeRMP2006,*KeimerNature2015, NormanNature1998, *ShenScience2005, *TanakaScience2006, *KanigelNatPhys2006, *KanigelPRL2007}.
Both gap and arcs are widely considered hallmarks of this underdoped region of the phase diagram, and have drawn significant attention aimed at uncovering their origin(s) \footnote{The word pseudogap has been variously used to describe both spectral gap and Fermi arc phenomena; here we note the distinction between these phenomena.}.  However, at a doping near optimal superconductivity, the Fermi arcs undergo an abrupt transition to a ``large" pocket consistent with a conventional area proportional to $1+p$ \cite{PlatePRL2005, ProustNature2008, BadouxNature2016, CollignonPRB2017}. A crucial challenge remains to identify the appropriate ground state(s) that underlie the theoretical framework on \textit{both} sides of this transition.

On the overdoped side, long thought to be a Fermi liquid, several recent reports of anomalous behavior call into question the conventional interpretation \cite{LegrosArXiv2018,PlatePRL2005,CooperScience2009,BozovicNature2016,BozovicNature2017,ArmitageArXiv2018}. 
The observations of resistivity linear in temperature \cite{CooperScience2009,LegrosArXiv2018} in Bi-based and La-based families challenge the expectations of standard Fermi liquid theory, and in MBE-grown La$_{2-x}$Sr$_x$CuO$_4$ compounds, there are reports of mysterious symmetry breaking \cite{BozovicNature2017} and anomalous scaling of the superfluid density with critical temperature \cite{BozovicNature2016, ArmitageArXiv2018}, although the latter remains controversial \cite{Lee-HonePRB2017, Lee-HoneArXiv2018}.  Furthermore, resonant inelastic x-ray experiments on \mbox{Tl-,} Y- \cite{LeTaconPRB2013}, and La-based \cite{DeanNatMater2013} compounds revealing spin fluctuations have been interpreted in terms of significant electron correlations, and there are now theoretical proposals~\cite{MrossPRL2012, MrossPRB2012} for how overdoped compounds may retain certain characteristic features of Fermi-liquid like behavior, while exhibiting fractionalization in the presence of strong stripe fluctuations.
Additional experiments are necessary to understand the overdoped compounds, and in particular to clarify the extent to which the effects of strong correlations may persist through the FS transition.

\begin{figure}[t]
	\centering
	\includegraphics[width=8.6cm]{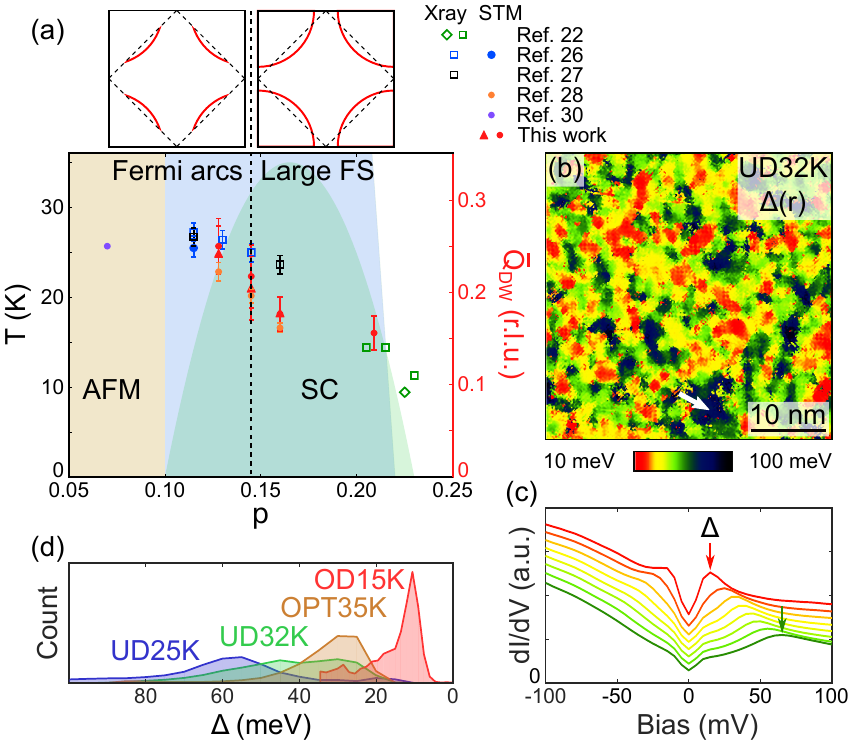}
	\caption{\label{f:phase_diagram} 
	    \textbf{ Using local electronic inhomogeneity and charge modulations to probe the Bi2201 phase diagram.} 
	    \textbf{(a)}~Phase diagram for Bi2201 in the region of the superconducting dome (green shading).  
	    With increasing $p$ at low temperature, the appearance of sharp antinodal quasiparticles (insets), indicative of a large normal state Fermi surface, occurs just below optimal doping (dashed line \cite{HeScience2014}), while the spectral gap persists into overdoped compounds (blue shading, with boundary marking the closing of the gap, as measured by ARPES \cite{KondoNatPhys2011} and NMR \cite{ZhengPRL2005}).  The data points mark existing sample-average measurements of the charge modulation wavevector, $\Qdwav$, in Bi2201 from x-ray scattering (green \cite{PengNatMater2017}, blue \cite{CominScience2014} and black \cite{PengPRB2016} open symbols) and STM (blue \cite{CominScience2014}, orange \cite{WiseNatPhys2008} and purple \cite{CaiNatPhys2016} filled symbols).  Red triangles and circles are the sample-average measurements of the $d$-form factor charge modulations in the $x$ and $y$ directions, respectively, from this work, with $p$ determined from Ando's conversion \cite{AndoPRB2000}, as described in the Supplemental Material.  The green squares and diamond are the annealed and as-grown samples, respectively, from Peng \textit{et al.}~\cite{PengNatMater2017}.  
	    \textbf{(b)}~The local spectral gap $\gap$ (shown for UD32K), measured as the local minimum in the second derivative of the empty state differential conductance spectrum measured at each point.  The white arrow marks the same location as in Fig.~\ref{f:fs-trans}h to highlight a region of large $\gap$ contributing to the Fermi arc QPI in Fig.~\ref{f:fs-trans}f.
	    \textbf{(c)}~Differential conductance spectra from UD32K, averaged over spatial regions binned by $\gap$ and offset vertically for clarity. 
	    \textbf{(d)}~Distributions of $\gap$ within the UD25K, UD32K, OPT35K and OD15K samples.  The histograms are normalized to have equal areas.
	    }
	\end{figure}

In nearly all cuprate families, charge order in the form of disordered charge modulations have been reported in underdoped compounds, with detection terminating at \cite{FujitaScience2014} or before \cite{BadouxPRX2016,BadouxNature2016} the doping where the FS transition occurs.
In \bsco{}, Bi2201, however, charge modulations extend into the overdoped regime \cite{HeScience2014,PengNatMater2017}.  These modulations, reflecting an ordering instability of the electronic system, therefore serve as a doping-dependent fingerprint of underlying electronic interactions, not just in the underdoped regime, but across the FS transition (Fig.\ \ref{f:phase_diagram}a). 
In the Bi2201 phase diagram, the FS transition occurs just below optimal doping
\footnote{To the authors' knowledge, the FS transition in Bi2201, near $p=0.14$, has been reported only by scanning tunneling microscopy~\cite{HeScience2014}, but Kondo \textit{et al.}'s ARPES measurements (compare Figs. 3e and f of Ref.\ \onlinecite{KondoNature2009}) also show that the spectral weight in the sharp quasiparticle peak at the antinode vanishes between the optimally doped and underdoped UD23K compounds.}, 
while the spectral gap persists in the presence of the large FS \cite{HeScience2014, KondoNatPhys2011, ZhengPRL2005}.
Peng \textit{et al.} \cite{PengNatMater2017} recently investigated the charge modulation structure across the closing of the gap in the far-overdoped region, but it remains crucial to clearly define how the wavevector, $\Qdw$, evolves across the FS transition (Fig.\ \ref{f:phase_diagram}a).  Existing measurements in this doping range \cite{CominScience2014, PengPRB2016, WiseNatPhys2008}, exhibit large scatter and doping coverage insufficient to clearly establish the $\Qdw$ trend on the underdoped side.  Furthermore, with the exception of Ref.\ \onlinecite{WiseNatPhys2009}, measurements have not taken into account the electronic inhomogeneity within these samples, even though the influence of annealing suggests that the bulk-averaged $\Qdwav$ is sensitive to disorder \cite{PengNatMater2017}.  Specifically, examining $\Qdwav$ in Fig.~\ref{f:phase_diagram}a, it is unclear (1) if a single doping-dependent incommensurate trend should be drawn through all of the measurements from $p=$ 0.11 to 0.23 and (2) how the charge modulation structure in this range relates to the commensurate modulation observed in a lightly doped compound near the insulating state \cite{CaiNatPhys2016}.

Here, we use the spatial dependence of the Bi2201 density wave (DW) as a probe of the parent states in both Fermi arc and large FS regions. We find that the FS transition marks the boundary between two distinct ground states that give rise to commensurate and incommensurate charge modulations, respectively. Furthermore, concurrent mapping of the DW and FS demonstrates that conventional Fermiology is insufficient to explain the overdoped evolution of $\Qdw$.


\begin{figure*}
	\centering
	\includegraphics[width=17.8cm]{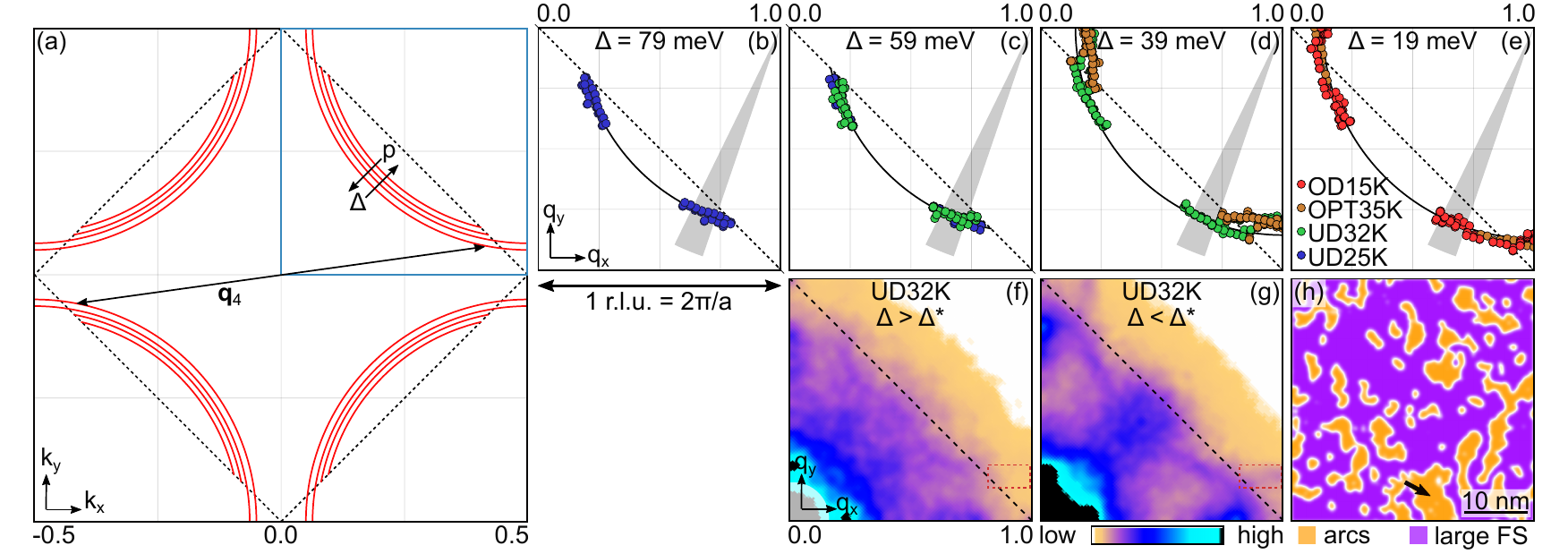}
	\caption{\label{f:fs-trans} 
	    \textbf{ A continuous doping axis from local electronic inhomogeneity.} 
	    \textbf{(a)} Fermi surface evolution with local doping, over the range of $\gap$ and $p$ studied here.  The double-headed arrow indicates an example $\myvec{q_4}$ scattering vector.  The blue box marks the quadrant that is shown for the QPI panels.   
	    \textbf{(b--e)} The QPI evolution with the local gap is shown by data points marking the peak positions extracted from gap-masked $Z(\myvec{q})$ for all four samples.  Refer to Appendix~\ref{ap:FS-from-QPI} and Supplemental Material Fig.~S1.  On each panel, data is compiled from masked regions with average $\gap$ within a 4~meV range, centered at the indicated value.  The black lines are circles determined by the average radius of the near-nodal data within a fixed angular range indicated by gray shading. For $\gap > \gapc$,
	    points are extracted only for regions inside the dashed line, as the intensity of antinodal QPI is negligible.  
	    \textbf{(f)} UD32K QPI from masking $Z(\myvec{r})$ by the yellow regions in (h), with $\gap$ primarily greater than $\gapc$.  QPI from near-nodal quasiparticles is strong but antinodal QPI (red dashed box) is not visible. 
	    \textbf{(g)} UD32K QPI from masking $Z(\myvec{r})$ by the purple regions in (h), with $\gap$ primarily less than $\gapc$.  QPI extends out to $q_{x,y}=1$~r.l.u. (red dashed box), indicating the presence of antinodal quasiparticles arising out of the antinodal normal state Fermi surface.  The gap-masked $Z(\myvec{q})$ in (f) and (g) is integrated from 10~meV to 25~meV, fourfold symmetrized, and smoothed with a Gaussian filter of width 0.015~r.l.u.\ to reduce the appearance of noise.
	    \textbf{(h)} Spatial division of UD32K into regions with $\gap$ primarily greater than (yellow) or less than (purple) $\gapc$ (Appendix~\ref{ap:ud32-qpi}).
	    The arrow in (h) marks the same location as in Fig.~\ref{f:phase_diagram}b to highlight a region of large $\gap$ contributing to the Fermi arc QPI.
	    All $\myvec{k}$ and $\myvec{q}$ axes are in reciprocal lattice units (1~r.l.u.\ = $2\pi/a$).  
	    }
	\end{figure*}

\section{Inhomogeneous Fermi Surface Transition}

Within a single Bi-based cuprate crystal, an average doping of $p$ holes per unit cell produces a highly inhomogeneous spatial distribution, resulting in large variations in the local electronic properties~\cite{ZeljkovicScience2012, WiseNatPhys2009, McElroyPRL2005, FeiArXiv2018, McElroyScience2005, PiriouNatCom2011, KinodaPRB2005}.  
In a scanning tunneling microscope, the local electronic density of states is typically measured by the spatially resolved differential conductance, $g(\myvec{r},E=eV) \equiv dI/dV(\myvec{r},V)$, where $V$ is the sample bias and $I(\myvec{r},V)$ is the tunneling current.  Binning and averaging the local spectra by gap size, $\gap(\myvec{r})$, as shown in Figs.\ \ref{f:phase_diagram}b and c, demonstrates the variation of the spectrum over a large field of view.  Previous work has shown that smaller $\gap$ corresponds to higher hole concentration, both locally within each sample \cite{ZeljkovicScience2012, FeiArXiv2018, McElroyScience2005} and globally from sample to sample \cite{MiyakawaPRL1998, *WhitePRB1996, *HarrisPRB1996, *DingPRL2001, *HufnerRPP2008}.  Thus, the same spectrum can be found locally in samples with different global $p$, and the overlapping $\gap$ distributions from the four samples studied in this work, UD25K, UD32K, OPT35K and OD15K (Fig.~\ref{f:phase_diagram}d), allow us to move continuously from underdoped (UD) to overdoped (OD) in the phase diagram, using spatial masking \cite{WiseNatPhys2009} to hone in on a single local doping within a larger field of view (Appendix~\ref{ap:gap-masking}).

To validate the use of local doping to construct the Bi2201 phase diagram, we demonstrate that the evolution of the FS with $\gap$ mimics that of bulk samples with $p$.  We calculate the ratio map $Z(\mathbf{r},E) \equiv g(\mathbf{r},E)/g(\mathbf{r},-E)$, which enhances Bogoliubov quasiparticle interference (QPI) and eliminates artifacts associated with the tip-sample junction setup~\cite{FujitaJPCS2008}. The normal state FS can be inferred from QPI in the superconducting state as follows \cite{WangPRB2003,HoffmanScience2002, McElroyNature2003}.  The QPI signal is dominated by wavevectors connecting regions of high density of states, i.e.\ extrema in the Bogoliubov dispersion $E(k) = \pm \sqrt{\epsilon_k^2 + |\Delta_k|^2}$, where $\epsilon_k$ is the normal state band dispersion, and $\Delta_k$ is the momentum dependent superconducting gap.  For a given angle, extrema in $E(k)$ are given by $\epsilon_k = E_F$ (normal state Fermi energy); thus, the Bogoliubov QPI dispersion traces out the normal state FS.  In particular, it is well-established that the $\myvec{q_4}(E)$ channel (Fig.~\ref{f:fs-trans}a) traces out $2k_F$ \cite{McElroyNature2003, HeScience2014, FujitaScience2014}, in excellent agreement with the normal state FS measured by ARPES \cite{McElroyNature2003, HanaguriNatPhys2007}.   
The $\myvec{q_4}$ wavevectors---extracted as a function of $\Delta$ by selecting a range of $\Delta$ values in $\Delta(\myvec{r})$ to mask the $Z(\myvec{r})$ data (Appendix~\ref{ap:gap-masking})---together describe a single evolution of the momentum-space electronic structure extending across all samples (Fig.~\ref{f:fs-trans}a--e).  Regions with a small gap ($\gap=19$~meV, Fig.~\ref{f:fs-trans}e) exhibit QPI tracing out a large normal state FS: the Bogoliubov quasiparticles near the antinodes at the edge of the Brillouin zone, $k_{x,y} = \pm\pi/a$ (Fig.~\ref{f:fs-trans}a), generate scattering with $\myvec{q_4}$ wavevectors that extend  out to $q_{x,y}=\pm2\pi/a$.  Moving to larger $\gap$ (Fig.~\ref{f:fs-trans}d), the $\myvec{q_4}$ trajectory shrinks, consistent with decreasing hole concentration, and the full evolution of the FS size inferred from the QPI (Supplemental Material Fig.\ S3) confirms that $\gap$ is well correlated to doping, in agreement with previous observations~\cite{PiriouNatCom2011,ZeljkovicScience2012,WiseNatPhys2009}.  

For larger $\gap$, the absence of observable QPI near $q_{x,y}=\pm2\pi/a$ (Supplemental Material Fig.\ S1) is consistent with a normal state Fermi arc that lacks sharp antinodal (AN) quasiparticles \cite{HeScience2014,FujitaScience2014}.  Quantitatively, the intensity of the AN QPI in the UD32K sample (Fig.~\ref{f:dw-qpt}a) decreases with increasing $\gap$ before settling at a constant value indistinguishable from the background, indicating the transition occurs at $\gap \approx 50$~meV, which we label $\gapc$.
Spatially dividing the data into regions of $\gap>\gapc$ and $<\gapc$ (Figs.\ \ref{f:fs-trans}f and g) shows that QPI associated with a large FS and Fermi arcs both exist within the sample, such that one can move on the phase diagram from one side of the transition to the other spatially.  This spatial division emphasizes that the electronic structure is determined on length scales similar to the $\gap$ correlation length.  The FS $p$ evolution of the cuprate phase diagram is therefore reproduced locally as a function of $\gap$, with the transition from Fermi arcs to large FS occurring at $\gapc\approx 50$~meV. 

\begin{figure}
	\centering
	\includegraphics[width=8.6cm]{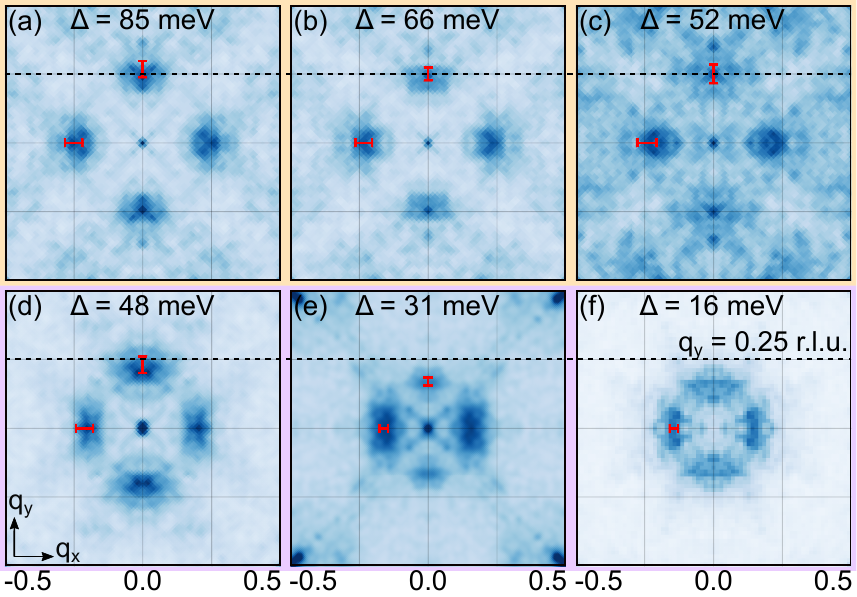}
	\caption{\label{f:dw-evolution} 
	    \textbf{ $\mathbf d$-form factor charge modulations.}  Examples of gap-masked $D(\myvec{q})$ where the indicated $\gap$ is the average value within each masked area.  The data are symmetrized along the $q_x$ and $q_y$ directions, and Gaussian smoothed with a 0.01~r.l.u.\ width.  
	    \textbf{(a--c)} Data with $\gap > \gapc$ (Fermi arc regime).
	    \textbf{(d--f)} Data with $\gap < \gapc$ (large FS regime).  
	    The thin gray lines have a spacing of 0.25 r.l.u.\ to facilitate comparison to the Fermiology in Fig.~\ref{f:fs-trans}.  The red symbols mark $\Qdw$ determined for $x$ and $y$ modulation directions, where the bar length indicates $\pm$ the estimated standard deviation of spatial fluctuations within each masked area. No measurement of $\Qdw$ is made for $q_y$ modulations in OD15K (f), as the wavevector is not sufficiently well defined.  The panels correspond to data from UD25K (a--c), UD32K (d), OPT35K (e) and OD15K (f).
	    }
	\end{figure}

\section{Commensurate to Incommensurate Transition}

We now determine the doping dependence of the charge modulation wavevector $\Qdw$ as a function of $\gap$ to look for signatures of the change in ground state at $\gapc$.  We examine $D(\myvec{r})$, the $d$-form factor ($d$FF) component~\cite{FujitaPNAS2014} of 
$\sum_{0<\epsilon<E} g(\mathbf{r},\epsilon) / \sum_{-E < \epsilon <0} g(\mathbf{r},\epsilon) \sim I(\mathbf{r},E) / I(\mathbf{r},-E)$, 
where the integration over an energy range larger than the typical $\gap$ for each sample enhances the DW signal~\cite{HamidianNatPhys2016}.  In all four samples, the amplitude of the Fourier transform $D(\myvec{q})$ has broad peaks at $(\pm\Qdw,0)$ and $(0,\pm\Qdw)$, near the charge modulation wavevectors that have been observed by previous experiments, indicating clearly that a $d$FF DW exists in all (Supplemental Material Fig.~S5).  We employ the demodulation phase residue minimization technique of Mesaros \textit{et al.}~\cite{MesarosPNAS2016} to make $\Qdw$ measurements robust against the strong disorder apparent from the broad shape of the DW peaks. 

Comparing Figs.~\ref{f:dw-evolution} a--c to d--f reveals that the $d$FF DW exhibits distinct evolutions for small and large spectral gap regions. From 16 meV to 48 meV, $\Qdw$ increases from 0.15 to 0.23 r.l.u., matching the change in wavevector that has been measured by resonant x-ray techniques from $p \approx 0.20$ to near-optimal doping \cite{PengNatMater2017,CominScience2014,PengPRB2016} and consistent with an evolving incommensurate wavevector.  However from $\gap \approx 50$ meV to $\gap \approx 85$~meV, no significant increase is observed (Fig.~\ref{f:dw-evolution}a--c). 
The constant value of $\Qdw$ near 0.25~r.l.u.\ suggests a dominant commensurate instability.  In fact, the entire $\Qdw(\gap)$ evolution (Fig.~\ref{f:dw-qpt}b) is consistent with a commensurate to incommensurate transition at a location indistinguishable from $\gapc$.  The dashed line showing the expected broadening of the underlying commensurate (yellow) to incommensurate (purple) trend describes the data accurately (Supplemental Material Sec.~SX).
Furthermore, the coincident changes in Fermiology and DW commensurability strongly suggest the presence of a quantum phase transition at $\gapc$. 

\subsection{Commensurate density wave}  
The observed wavevector in the Fermi arc state ($\gap>\gapc$) is consistent with a commensurate four-unit-cell charge modulation.  The average $\Qdwav$ from UD25K, which lies almost entirely on the underdoped side of the transition, is 0.25(3)~r.l.u.\ and 0.24(3)~r.l.u.\ for $x$ and $y$ directions, respectively, where the errors represent  the estimated standard deviation of spatial variations across the entire field of view.   
Resonant x-ray experiments \cite{CominScience2014} have reported a doping-dependent $\Qdwav$ in this same doping range,  down to $p=0.115$.  However, this apparent discrepancy can be understood by considering: (1) the FS transition occurs near $p=0.14$ \cite{HeScience2014} and beyond this point, we also observe an incommensurate wavevector, (2) the local inhomogeneity could plausibly induce a doping dependence of $\Qdwav$ even in samples with average $p$ below the transition, and (3) the strongly disordered structure of charge modulations limits the precision with which the value of $\Qdw$ (and $\Qdwav$) can be determined unambiguously~\cite{MesarosPNAS2016}.  For the samples studied here, ignoring local doping variations and taking the average $\Qdwav$ value for each sample hides the kink at the FS transition, and produces a trend of decreasing $\Qdwav$ with doping similar to previous reports (Fig.\ \ref{f:phase_diagram}a).  Furthermore, while our data do not rule out a small doping dependence, recent experiments on Bi2212 \cite{MesarosPNAS2016,ZhangArXiv2018, ZhaoNatMat2018} and very underdoped Bi2201 \cite{CaiNatPhys2016} are also consistent with a $Q=0.25$~r.l.u.\ commensurate DW within the Fermi arc regime arising from proximity to the Mott insulating state.

\subsection{Incommensurate density wave}  
\begin{figure}
	\centering
	\includegraphics[width=8.6cm]{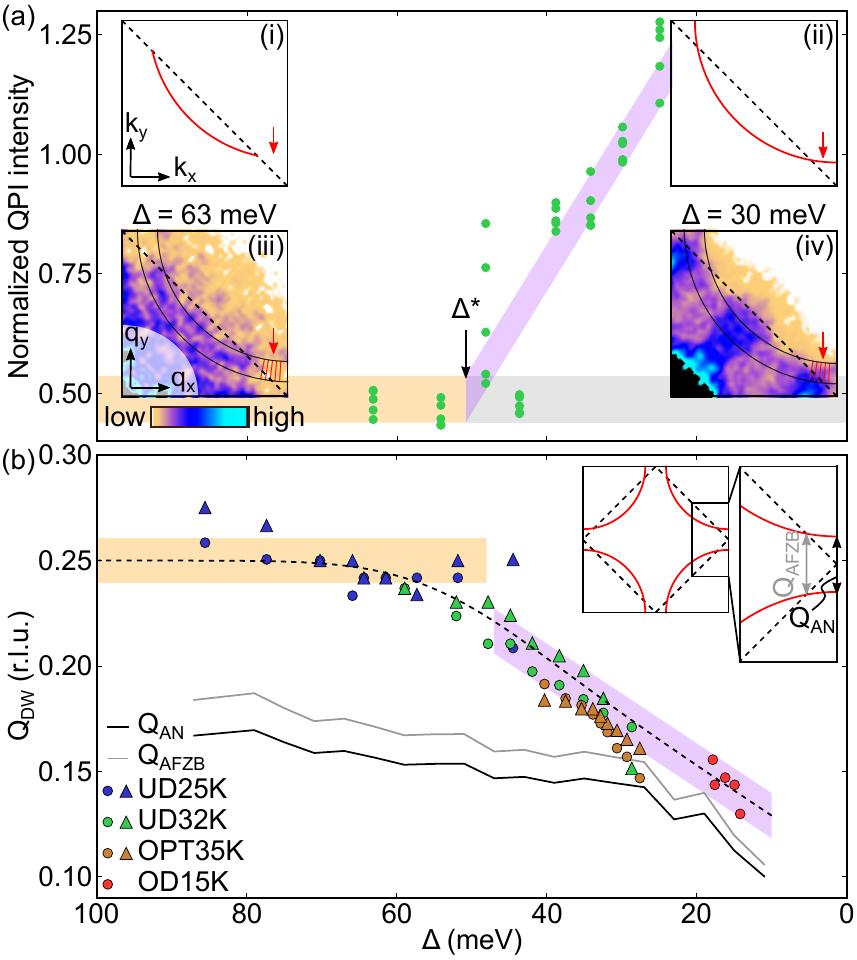}
	\caption{\label{f:dw-qpt} 
	    \textbf{ Simultaneous DW and FS transitions.} 
	    \textbf{(a)} Intensity of AN QPI in UD32K determined at five locations (red lines in the insets (iii, iv)) and normalized to the intensity of near-nodal QPI (Appendix~\ref{ap:qpi-intensity}).  $\gapc$ indicates the approximate location where the AN QPI becomes indistinguishable from the background noise.  The thick shaded lines are guides to the eye.  Insets (i,ii) show the approximate Fermi surface structure inferred from two examples of the gap-masked $Z(\myvec{q})$ (iii, iv) without and with AN QPI.  \textbf{(b)} Wavevector ($\Qdw$) of the $d$FF charge modulations in the $x$ (triangles) and $y$ (circles) directions, extracted from gap-masked $D(\myvec{r})$.
	    Refer to Supplemental Material Fig.~S9 for the standard deviation of $\Qdw$ for each value of $\gap$.
	    The thick shaded lines indicate the commensurate to incommensurate trend underlying the dashed line, which includes the expected effect of Gaussian smoothing ($\sigma=12$ meV) in $\Delta$ due to the resolution of the masking technique (Supplemental Material Fig.~S10). 
	    The gray and black lines indicate the Fermiology-driven candidate wavevectors, as indicated schematically in the insets, and based on the circular Fermi surface models shown in Figs.~\ref{f:fs-trans}a--e.
	    }
	\end{figure}

What is the mechanism for the incommensurate $\Qdw$ observed for $\gap < \gapc$?  Previous work interpreted the monotonically decreasing $\Qdw$ as evidence of a FS instability that follows the growing FS hole pocket \cite{WiseNatPhys2009}. In this picture, there are two natural candidates for $\Qdw$: (1) $\Qan$ that connects nested antinodal segments of the FS, and (2) $\Qafzb$ that connects the points where the FS crosses the antiferromagnetic zone boundary (AFZB), the hotspots for $(\pi,\pi)$ spin fluctuations.
Both $\Qan(\gap)$ and $\Qafzb(\gap)$ are shown in Fig.~\ref{f:dw-qpt}b. For the largest $p$ (smallest $\gap$), $\Qdw$ becomes similar to these Fermiology-derived wavevectors.  However, upon decreasing $p$ towards the transition, $\Qdw$ grows more rapidly than the FS evolves. This unexpected discrepancy between $\Qdw$ and Fermiology constitutes our second major finding.

\section{Discussion}

Whereas most recent theories for charge modulations in the cuprates have aimed to explain an incommensurate DW in the presence of Fermi arcs, here we are discussing (1) a \textit{commensurate} DW in the presence of Fermi arcs and (2) an incommensurate DW occurring in the presence of the \textit{large FS}.  This leads to two important distinctions.  First, in the Fermi arc regime, strong interactions are expected, and the associated renormalization can affect a hotspot (HS) wavevector, $\Qhs$.  For an instability of the large FS, however, it is not \textit{a priori} clear that there should be any influence of correlations on such a $\Qhs$. This first distinction reconciles our conclusion with Ref.~\cite{CominScience2014}'s interpretation that $\Qdwav$ is Fermiology driven~\footnote{Ref.~\cite{CominScience2014} concludes that $\Qdwav$ is explained by a FS instability, apparently in contradiction to the findings reported herein.  However, Ref.~\cite{CominScience2014} models an incommensurate $\Qdwav$ in the presence of Fermi arcs, where the renormalization associated with the arc phenomenology generates a $\Qhs$ significantly larger than $\Qafzb$}.  We note that additional factors, such as coupling to the lattice, may affect the observed $\Qdw$.  Second, theoretical studies have found that the dominant charge density wave (CDW) instability of the large FS in the presence of exchange interactions has a wavevector along the $(\pm q,\pm q)$ direction rather than $(0,\pm q)$ or $(\pm q,0)$, and the presence or absence of antinodal states is important in stabilizing the former or latter orientation, respectively~\cite{ChowdhuryPRB2014}.  It is therefore significant that the orientation of the charge modulations does not change at $\gapc$.

To search for information about the nature of the transition at $\gapc$, we consider that generically in CDW systems, a discontinuity or sharp jump in $\Qdw$ occurs at a commensurate to incommensurate transition~\cite{Bak1982RPP}.  From this work, we cannot distinguish between a continuous $\Qdw$ or one with a small jump, as shown by the yellow and purple trends in Fig.~\ref{f:dw-qpt}b (see also Supplemental Material Fig.~S10j).  To extract this information, $\gapc$ and the incommensurate $\Qdw(\gap)$ would need to be determined with reduced uncertainties.  However, if present, a continuous $\Qdw$, which generically requires fine tuning, would imply a mechanism-derived constraint on the position of $\gapc$ not explained by existing theoretical models of the Fermi surface transition. 

Ubiquity across underdoped compounds has widely been cited to motivate studying charge modulations as a route to understanding Fermi arc physics and the mechanism behind high-$T_c$ superconductivity in cuprates.  However, the relationship among these three phenomena has remained an open question.  The coincidence of the DW and FS transitions observed here establishes an intimate link between the DW and the presence of Fermi arcs, and furthermore suggests that the same interactions which generate the commensurate instability may also be responsible for the arc phenomenology.

In summary, we report three concrete observations: (1) There is a commensurate to incommensurate transition of $\Qdw$ at a doping consistent with that of the FS transition; (2) on the underdoped side, $\Qdw$ is consistent with a commensurate four-unit-cell modulation, and (3) on the overdoped side, the doping dependence of $\Qdw$ is stronger than that of the FS size. 

\section{Acknowledgments}
The authors thank Subir Sachdev, Patrick A.~Lee, Andrej Mesaros, B.~J.\ Ramshaw, Simon Verret, Milan Allan, J.~C.\ S\'eamus Davis, Makoto Hashimoto, Maude Lizaire, and Louis Taillefer for helpful discussions.  
This work was supported by the Gordon and Betty Moore Foundation's EPiQS Initiative through Grant No.\ GBMF4536 and the National Science Foundation under Grant No.\ DMR-1341286.  D.C.\ is supported by a postdoctoral fellowship from the Gordon and Betty Moore Foundation, under the EPiQS initiative, Grant No.\ GBMF-4303, at MIT.

\appendix

\section{STM data}
Differential conductance maps were collected  in scanning tunneling microscopes at 6 K with the following tip-sample junction setup conditions: 100 mV and 100 pA for OD15K, -100 mV and 400 pA for OPT35K, -200 mV and 400 pA for UD32K, and -150 mV and 400 pA for UD25K.  Measurements used a lock-in technique with 2 mV, 10 mV, 5 mV, and 5 mV bias modulations, respectively.
Data were corrected for artificial distortions due to instrument drift to register the simultaneously recorded topography to a perfect lattice \cite{LawlerNature2010}.

\section{Gap-masking technique}
\label{ap:gap-masking}
The gap map $\gap(\myvec{r})$ is calculated by finding the position of the local minimum in the second derivative of the empty state differential conductance for the spectrum $g(\myvec{r},E)$ at each pixel $\myvec{r}$, where Gaussian smoothing in energy reduces errors from noise in the data.  The gap masks are then generated by dividing the values of $\gap$ into bins with equal counts.  The mask for bin $b$, $M_b(\myvec{r})$, has a value of 1 if $\gap(\myvec{r})$ is in $b$, or zero otherwise.  In order to reduce periodic structure in the masks arising from the atomic corrugation or the charge modulations \cite{WiseNatPhys2009}, a bilateral filter is applied to $\gap(\myvec{r})$ before generating the masks.

To obtain clear images in momentum transfer, $\myvec{q}$, space, we use an additive masking technique, where for each bin $b$, we look for the change in Fourier transform amplitude when adding $b$ into the field of view.  The additive masks therefore include bins summed up to a bin $b$:
$M_{1,b}(\myvec{r}) = \sum_{j=1}^{b} M_j(\myvec{r})$.  
To avoid introducing artifacts from spatial structure of the masks, we apply Gaussian smoothing to the mask, with a spatial resolution of $1/w$.  The filtered masks $\mathcal{M}_{1,b}(\myvec{r}) = G_{w}(\myvec{r}) * M_{1,b}(\myvec{r})$ can now have any value in the range [0,1], where $G_{w}$ is the Gaussian filter, and $*$ indicates convolution.  This filtering imposes a spatial resolution of the masks and implies a spread in the $\gap$ distribution within each mask (Supplemental Material Fig.\ S10).

To explain the additive masking technique, we describe how the images in Fig.\ \ref{f:dw-evolution} are generated.  The additive masks are applied to the $D(\myvec{r})$ map in real space: 
$ D_{a,b}(\myvec{r}) = \mathcal{M}_{a,b}(\myvec{r}) D(\myvec{r}) $.  
To visualize the charge order peaks from bin $b$, we take the difference of the absolute values of the Fourier transforms, including and not including bin $b$.  The additive masking can either add up $\gap$ values from small to large (forward, fw) or in the reverse direction (backward, bw).  For these two cases, the gap-masked $D(\myvec{q})$ is
\begin{eqnarray}
    D^{\mathrm{fw}}_b(\myvec{q}) 
    &=& D_{1,b}(\myvec{q}) - D_{1,b-1}(\myvec{q}) \\
    D^{\mathrm{bw}}_b(\myvec{q}) 
    &=& D_{b,N}(\myvec{q}) - D_{b+1,N}(\myvec{q}),
    \end{eqnarray}
where $N$ is the total number of bins and $D_{a,b}(\myvec{q})$ is a real number, the amplitude of the Fourier transform.  Unless otherwise specified, functions of $\myvec{q}$ refer to amplitudes of Fourier transforms.  The panels of Fig.\ \ref{f:dw-evolution} have Gaussian smoothing applied with width 0.01~r.l.u., to reduce the appearance of noise.

The QPI shown in the insets of Fig.\ \ref{f:dw-qpt} are $Z_b^{\mathrm{fw}}(\myvec{q})$ generated from $Z(\myvec{r})$ using this same technique.  Additional gap-masked QPI from UD32K in Supplemental Material Fig.\ S1 shows the evolution of the Fermi surface structure across the full range of $\gap$ within the sample.  These QPI images have Gaussian smoothing applied with width 0.015~r.l.u., and have been fourfold symmetrized.

For analysis of the Fermi surface structure, we use overlapping bins, i.e. \begin{eqnarray}
    Z^{\mathrm{fw}}_b(\myvec{q}) 
    &=& Z_{1,b}(\myvec{q}) - Z_{1,b-n}(\myvec{q}) \\
    Z^{\mathrm{bw}}_b(\myvec{q}) 
    &=& Z_{b,N}(\myvec{q}) - Z_{b+n,N}(\myvec{q}),
    \end{eqnarray}
in order to measure the QPI wavevectors at finer spaced intervals of $\gap$ with more variations in the masks.  The integer $n$ determines the overlap between subsequent bins.

The average value of the gap for bin $b$ is then
\begin{eqnarray}
    \gap_b^{\mathrm{fw}} &=& \frac{\sum_{\myvec{r}} \gap(\myvec{r}) 
                    \left[ \mathcal{M}_{1,b}(\myvec{r}) 
                    - \mathcal{M}_{1,b-n}(\myvec{r}) \right]}
                    {\sum_{\myvec{r}} 
                    \left[ \mathcal{M}_{1,b}(\myvec{r}) 
                    - \mathcal{M}_{1,b-n}(\myvec{r}) \right]},\\
    \gap_b^{\mathrm{bw}} &=& \frac{\sum_{\myvec{r}} \gap(\myvec{r}) 
                    \left[ \mathcal{M}_{b,N}(\myvec{r}) 
                    - \mathcal{M}_{b+n,N}(\myvec{r}) \right]}
                    {\sum_{\myvec{r}} 
                    \left[ \mathcal{M}_{b,N}(\myvec{r}) 
                    - \mathcal{M}_{b+n,N}(\myvec{r}) \right]},
    \end{eqnarray}
where $n=1$ for distinct, as opposed to overlapping, bins.

Here, we use $N= 19$ for UD25K, 31 for UD32K, 31 for OPT35K, and 21 for OD15K, with $n=4$, and $w=0.20$~r.l.u.

For $D(\myvec{q})$ in Fig.\ \ref{f:dw-evolution}, we use $N=9$ with $n=1$ for UD25K, UD32K and OPT35K, and $N=21$ with $n=4$ for OD15K.  For OD15K, only the data for bins $b=$ 5, 9, 13, 17, and 21 are plotted in Fig.\ \ref{f:dw-qpt}b to present data from distinct bins, as are used on the other three samples.  The mask smoothing parameter $w$ was chosen to include the resolution $\Lambda$ of determining $\Qdw$ (Appendix~\ref{ap:residue}): $w=\Lambda=0.10$~r.l.u.\ for UD25K, $w=\Lambda=0.10$~r.l.u.\ for UD32K, $w=\Lambda=0.04$~r.l.u.\ for OPT35K and $w=\left(\Lambda^{-2}+0.20\,\mathrm{r.l.u.}^{-2}\right)^{-1/2}$ with $\Lambda=0.04$~r.l.u.\ for OD15K.

\section{Fermi surface structure from gap-masked QPI}
\label{ap:FS-from-QPI}

$Z(\myvec{q})$ was integrated over the low energy layers to capture the full dispersion of the Bogoliubov quasiparticles, from 1.5 meV to 9 meV for OD15K, 5 meV to 5 meV for OPT35K, 10 meV to 25 meV for UD32K, and 5 meV to 15 meV for UD25K.

For each $\gap$ bin, the QPI wavevectors, as shown in Fig.\ \ref{f:fs-trans}, are extracted from the positions of peaks in one-dimensional cuts through $Z_b^{\mathrm{fw}}(\myvec{q})$ and $Z_b^{\mathrm{bw}}(\myvec{q})$, and are shown in Supplemental Material Fig.\ S1 for UD32K.  To quantitatively determine the size of the Fermi surface, the wavevectors from all samples are binned together, as shown in Fig.\ \ref{f:fs-trans}.  A circular hole pocket is determined from the average radius of the data in the range $\theta = 0.105\pi$ to $0.145\pi$, where $\theta$ is defined in Supplemental Material Fig.\ S1, and the range is selected because this near-nodal QPI is consistently measured across the Fermi surface transition (unlike the antinodal QPI) and is least influenced by nearby scattering channels or the DW signal.  The evolution of the Fermi surface radius is shown in Supplemental Material Fig.\ S3 and explicitly demonstrates that $\gap$ tracks the local doping.

\section{Fermi arc and large Fermi surface QPI}
\label{ap:ud32-qpi}
To generate Figs.\ \ref{f:fs-trans}f and g, we locate the bin $b^*$ that has the largest average gap value below 49~meV, the average of $\gapc$ estimates based on forward and backward masked data shown in Fig.~\ref{f:dw-qpt}a and Supplemental Material Fig.~S4b.  Bins $b\le b^*$ primarily have $\gap < \gapc$, and $b>b^*$ primarily $\gap>\gapc$.  This division is only approximate due to the Gaussian smoothing of the masks (Supplemental Material Fig.\ S10).  To probe QPI from the two regimes divided by $\gapc$, Figs.~\ref{f:fs-trans}f and g show $Z_{b^*+1,N}(\myvec{q})$, and $Z_{1,b^*}(\myvec{q})$, respectively.  Note that in this case, there is no subtraction after applying the Fourier transform.  Fig.\ \ref{f:fs-trans}h shows $\mathcal{M}_{1,b^*}(\myvec{r})$, where the colorscale interpolates between 0.0 (yellow) and 1.0 (purple).  Because the masks used to obtain f and g are related by $\mathcal{M}_{b^*+1,N}(\myvec{q}) = 1 - \mathcal{M}_{1,b^*}(\myvec{q})$, the yellow and purple indicate the regions primarily contributing to the Fermi arc and large Fermi surface QPI, respectively.

\section{Intensity of AN QPI in UD32K}
\label{ap:qpi-intensity}
In order to quantify the disappearance of the antinodal (AN) QPI, which is apparent directly in the data (Supplemental Material Fig.\ S1), we measure the intensity at cuts spaced at regular angular intervals, as shown in Supplemental Material Fig.\ S4.  The cuts are averaged over a transverse width of 0.07~r.l.u.\ and a length determined by the QPI radius (twice the FS radius) $\pm$ 0.06~r.l.u.  To compare QPI intensities from different bins, the intensities of the cuts are normalized by the average intensity of five cuts closer to the nodal QPI (dashed red lines in Supplemetal Material Fig.\ S4a).  Fig.\ \ref{f:dw-qpt}a tracks this normalized intensity for five cuts near the antinode (red lines in Fig.\ \ref{f:dw-qpt}a-iii and -iv).

\section{Extracting \texorpdfstring{$\bf \Qdw$}{Qdw} from $d$FF charge modulation}
\label{ap:residue}

We follow the procedure described by Mesaros \textit{et al.}~\cite{MesarosPNAS2016}, where $\Qdw$ is determined as the wavevector which minimizes the demodulation residue,
$R_\myvec{Q} = \sqrt{|R^x_\myvec{Q}|^2 +|R^y_\myvec{Q}|^2}$, 
over the field of view.  In the case of a strongly disordered density wave, this measurement has a more clearly defined interpretation than fitting peaks of the Fourier transformed data.  Both techniques are compared in Sec.\ SVIII of the Supplemental Material.  A detailed explanation of demodulation residue can be found in Ref.~\onlinecite{MesarosPNAS2016}.  In this section, we extend the technique for application to masked regions of the data's field of view.

$D(\myvec{r})$ is demodulated by the reference wavevector $\myvec{Q}$ in 
\begin{equation}
    \tilde{\Psi}_{\myvec{Q}}(\myvec{q}) = 
    \exp\left( \frac{- q^2}{2 \Lambda^2} \right) \tilde{\psi}(\myvec{q}+\myvec{Q}), \\    
    \end{equation}
where $\tilde{\psi}(\myvec{q})=\tilde{D}(\myvec{q})$ over a domain which isolates the charge order peak, $\tilde{D}$ is the complex-valued Fourier transform of $D(\myvec{r})$, and $\tilde{\Psi}_{\myvec{Q}}$, $\tilde{\psi}$ are complex-valued functions.  The Gaussian cutoff imposes a spatial resolution of $1/\Lambda$.

The $\Qdw$ measurement proceeds in each gap bin $b$ by integrating the residue only over the masked region:
\begin{eqnarray}
    R^\alpha_{b,\myvec{Q}}[\Psi] &=& 
        \int d^2\myvec{r} \, \mathcal{M}_b(\myvec{r}) \, 
        \operatorname{Re}\left[ \Psi^*_\myvec{Q}  (-i\partial_\alpha)  \Psi_\myvec{Q} \right], \nonumber \\
    \sigma^2 &=& 
        \frac{ 
            \int d^2\myvec{r} \, \mathcal{M}_b(\myvec{r}) \, \sum_\alpha
            \left[ \Psi^*_\myvec{Q}  (-i\partial_\alpha)  \Psi_\myvec{Q} \right]^2
        }{
            \int d^2\myvec{r} \, \mathcal{M}_b(\myvec{r}) \, 
            \left| \Psi_\myvec{Q} \right|^2
        },
    \end{eqnarray}
where $\alpha$ is either $x$ or $y$, $\Psi_\myvec{Q}(\myvec{r})$ is the inverse Fourier transform of $\tilde{\Psi}_\myvec{Q}(\myvec{q})$, and with $\myvec{Q}=\myvec{Q}_\mathrm{DW}$, $\sigma$ estimates the standard deviation of spatial fluctuations in the modulation wavevector.

For this analysis, we use $w=0.0$ as the mask smoothing parameter for UD25K, UD32K and OPT35K, with $\Lambda=0.10$, 0.10, and 0.04~r.l.u., respectively.  For OD15K, we use $w=0.20$~r.l.u.\ and $\Lambda=0.04$~r.l.u.  The samples with smaller wavevector require a smaller $\Lambda$ for $\Qdw$ to be robust against the choice of $\Lambda$.

\bibliography{refs}

\begin{thebibliography}{62}%
\makeatletter
\providecommand \@ifxundefined [1]{%
 \@ifx{#1\undefined}
}%
\providecommand \@ifnum [1]{%
 \ifnum #1\expandafter \@firstoftwo
 \else \expandafter \@secondoftwo
 \fi
}%
\providecommand \@ifx [1]{%
 \ifx #1\expandafter \@firstoftwo
 \else \expandafter \@secondoftwo
 \fi
}%
\providecommand \natexlab [1]{#1}%
\providecommand \enquote  [1]{``#1''}%
\providecommand \bibnamefont  [1]{#1}%
\providecommand \bibfnamefont [1]{#1}%
\providecommand \citenamefont [1]{#1}%
\providecommand \href@noop [0]{\@secondoftwo}%
\providecommand \href [0]{\begingroup \@sanitize@url \@href}%
\providecommand \@href[1]{\@@startlink{#1}\@@href}%
\providecommand \@@href[1]{\endgroup#1\@@endlink}%
\providecommand \@sanitize@url [0]{\catcode `\\12\catcode `\$12\catcode
  `\&12\catcode `\#12\catcode `\^12\catcode `\_12\catcode `\%12\relax}%
\providecommand \@@startlink[1]{}%
\providecommand \@@endlink[0]{}%
\providecommand \url  [0]{\begingroup\@sanitize@url \@url }%
\providecommand \@url [1]{\endgroup\@href {#1}{\urlprefix }}%
\providecommand \urlprefix  [0]{URL }%
\providecommand \Eprint [0]{\href }%
\providecommand \doibase [0]{http://dx.doi.org/}%
\providecommand \selectlanguage [0]{\@gobble}%
\providecommand \bibinfo  [0]{\@secondoftwo}%
\providecommand \bibfield  [0]{\@secondoftwo}%
\providecommand \translation [1]{[#1]}%
\providecommand \BibitemOpen [0]{}%
\providecommand \bibitemStop [0]{}%
\providecommand \bibitemNoStop [0]{.\EOS\space}%
\providecommand \EOS [0]{\spacefactor3000\relax}%
\providecommand \BibitemShut  [1]{\csname bibitem#1\endcsname}%
\let\auto@bib@innerbib\@empty
\bibitem [{\citenamefont {Lee}\ \emph {et~al.}(2006)\citenamefont {Lee},
  \citenamefont {Nagaosa},\ and\ \citenamefont {Wen}}]{LeeRMP2006}%
  \BibitemOpen
  \bibfield  {author} {\bibinfo {author} {\bibfnamefont {P.~A.}\ \bibnamefont
  {Lee}}, \bibinfo {author} {\bibfnamefont {N.}~\bibnamefont {Nagaosa}}, \ and\
  \bibinfo {author} {\bibfnamefont {X.-G.}\ \bibnamefont {Wen}},\ }\bibfield
  {title} {\enquote {\bibinfo {title} {Doping a {Mott} insulator: Physics of
  high-temperature superconductivity},}\ }\href {\doibase
  10.1103/RevModPhys.78.17} {\bibfield  {journal} {\bibinfo  {journal} {Rev.
  Mod. Phys.}\ }\textbf {\bibinfo {volume} {78}},\ \bibinfo {pages} {17--85}
  (\bibinfo {year} {2006})}\BibitemShut {NoStop}%
\bibitem [{\citenamefont {Keimer}\ \emph {et~al.}(2015)\citenamefont {Keimer},
  \citenamefont {Kivelson}, \citenamefont {Norman}, \citenamefont {Uchida},\
  and\ \citenamefont {Zaanen}}]{KeimerNature2015}%
  \BibitemOpen
  \bibfield  {author} {\bibinfo {author} {\bibfnamefont {B.}~\bibnamefont
  {Keimer}}, \bibinfo {author} {\bibfnamefont {S.~A.}\ \bibnamefont
  {Kivelson}}, \bibinfo {author} {\bibfnamefont {M.~R.}\ \bibnamefont
  {Norman}}, \bibinfo {author} {\bibfnamefont {S.}~\bibnamefont {Uchida}}, \
  and\ \bibinfo {author} {\bibfnamefont {J.}~\bibnamefont {Zaanen}},\
  }\bibfield  {title} {\enquote {\bibinfo {title} {From quantum matter to
  high-temperature superconductivity in copper oxides},}\ }\href {\doibase
  10.1038/nature14165} {\bibfield  {journal} {\bibinfo  {journal} {Nature}\
  }\textbf {\bibinfo {volume} {518}},\ \bibinfo {pages} {179--186} (\bibinfo
  {year} {2015})}\BibitemShut {NoStop}%
\bibitem [{\citenamefont {Norman}\ \emph {et~al.}(1998)\citenamefont {Norman},
  \citenamefont {Ding}, \citenamefont {Randeria}, \citenamefont {Campuzano},
  \citenamefont {Yokoya}, \citenamefont {Takeuchi}, \citenamefont {Takahashi},
  \citenamefont {Mochiku}, \citenamefont {Kadowaki}, \citenamefont
  {Guptasarma},\ and\ \citenamefont {Hinks}}]{NormanNature1998}%
  \BibitemOpen
  \bibfield  {author} {\bibinfo {author} {\bibfnamefont {M.~R.}\ \bibnamefont
  {Norman}}, \bibinfo {author} {\bibfnamefont {H.}~\bibnamefont {Ding}},
  \bibinfo {author} {\bibfnamefont {M.}~\bibnamefont {Randeria}}, \bibinfo
  {author} {\bibfnamefont {J.~C.}\ \bibnamefont {Campuzano}}, \bibinfo {author}
  {\bibfnamefont {T.}~\bibnamefont {Yokoya}}, \bibinfo {author} {\bibfnamefont
  {T.}~\bibnamefont {Takeuchi}}, \bibinfo {author} {\bibfnamefont
  {T.}~\bibnamefont {Takahashi}}, \bibinfo {author} {\bibfnamefont
  {T.}~\bibnamefont {Mochiku}}, \bibinfo {author} {\bibfnamefont
  {K.}~\bibnamefont {Kadowaki}}, \bibinfo {author} {\bibfnamefont
  {P.}~\bibnamefont {Guptasarma}}, \ and\ \bibinfo {author} {\bibfnamefont
  {D.~G.}\ \bibnamefont {Hinks}},\ }\bibfield  {title} {\enquote {\bibinfo
  {title} {Destruction of the {Fermi} surface in underdoped high-{$T_c$}
  superconductors},}\ }\href {\doibase 10.1038/32366} {\bibfield  {journal}
  {\bibinfo  {journal} {Nature}\ }\textbf {\bibinfo {volume} {392}},\ \bibinfo
  {pages} {157--160} (\bibinfo {year} {1998})}\BibitemShut {NoStop}%
\bibitem [{\citenamefont {Shen}\ \emph {et~al.}(2005)\citenamefont {Shen},
  \citenamefont {Ronning}, \citenamefont {Lu}, \citenamefont {Baumberger},
  \citenamefont {Ingle}, \citenamefont {Lee}, \citenamefont {Meevasana},
  \citenamefont {Kohsaka}, \citenamefont {Azuma}, \citenamefont {Takano},
  \citenamefont {Takagi},\ and\ \citenamefont {Shen}}]{ShenScience2005}%
  \BibitemOpen
  \bibfield  {author} {\bibinfo {author} {\bibfnamefont {K.~M.}\ \bibnamefont
  {Shen}}, \bibinfo {author} {\bibfnamefont {F.}~\bibnamefont {Ronning}},
  \bibinfo {author} {\bibfnamefont {D.~H.}\ \bibnamefont {Lu}}, \bibinfo
  {author} {\bibfnamefont {F.}~\bibnamefont {Baumberger}}, \bibinfo {author}
  {\bibfnamefont {N.~J.~C.}\ \bibnamefont {Ingle}}, \bibinfo {author}
  {\bibfnamefont {W.~S.}\ \bibnamefont {Lee}}, \bibinfo {author} {\bibfnamefont
  {W.}~\bibnamefont {Meevasana}}, \bibinfo {author} {\bibfnamefont
  {Y.}~\bibnamefont {Kohsaka}}, \bibinfo {author} {\bibfnamefont
  {M.}~\bibnamefont {Azuma}}, \bibinfo {author} {\bibfnamefont
  {M.}~\bibnamefont {Takano}}, \bibinfo {author} {\bibfnamefont
  {H.}~\bibnamefont {Takagi}}, \ and\ \bibinfo {author} {\bibfnamefont {Z.-X.}\
  \bibnamefont {Shen}},\ }\bibfield  {title} {\enquote {\bibinfo {title} {Nodal
  quasiparticles and antinodal charge ordering in
  {Ca$_{2-x}$Na$_x$CuO$_2$Cl$_2$}},}\ }\href {\doibase 10.1126/science.1103627}
  {\bibfield  {journal} {\bibinfo  {journal} {Science}\ }\textbf {\bibinfo
  {volume} {307}},\ \bibinfo {pages} {901--904} (\bibinfo {year}
  {2005})}\BibitemShut {NoStop}%
\bibitem [{\citenamefont {Tanaka}\ \emph {et~al.}(2006)\citenamefont {Tanaka},
  \citenamefont {Lee}, \citenamefont {Lu}, \citenamefont {Fujimori},
  \citenamefont {Fujii}, \citenamefont {Risdiana}, \citenamefont {Terasaki},
  \citenamefont {Scalapino}, \citenamefont {Devereaux}, \citenamefont
  {Hussain},\ and\ \citenamefont {Shen}}]{TanakaScience2006}%
  \BibitemOpen
  \bibfield  {author} {\bibinfo {author} {\bibfnamefont {K.}~\bibnamefont
  {Tanaka}}, \bibinfo {author} {\bibfnamefont {W.~S.}\ \bibnamefont {Lee}},
  \bibinfo {author} {\bibfnamefont {D.~H.}\ \bibnamefont {Lu}}, \bibinfo
  {author} {\bibfnamefont {A.}~\bibnamefont {Fujimori}}, \bibinfo {author}
  {\bibfnamefont {T.}~\bibnamefont {Fujii}}, \bibinfo {author} {\bibnamefont
  {Risdiana}}, \bibinfo {author} {\bibfnamefont {I.}~\bibnamefont {Terasaki}},
  \bibinfo {author} {\bibfnamefont {D.~J.}\ \bibnamefont {Scalapino}}, \bibinfo
  {author} {\bibfnamefont {T.~P.}\ \bibnamefont {Devereaux}}, \bibinfo {author}
  {\bibfnamefont {Z.}~\bibnamefont {Hussain}}, \ and\ \bibinfo {author}
  {\bibfnamefont {Z.-X.}\ \bibnamefont {Shen}},\ }\bibfield  {title} {\enquote
  {\bibinfo {title} {Distinct {Fermi}-momentum-dependent energy gaps in deeply
  underdoped {Bi2212}},}\ }\href {\doibase 10.1126/science.1133411} {\bibfield
  {journal} {\bibinfo  {journal} {Science}\ }\textbf {\bibinfo {volume}
  {314}},\ \bibinfo {pages} {1910--1913} (\bibinfo {year} {2006})}\BibitemShut
  {NoStop}%
\bibitem [{\citenamefont {Kanigel}\ \emph {et~al.}(2006)\citenamefont
  {Kanigel}, \citenamefont {Norman}, \citenamefont {Randeria}, \citenamefont
  {Chatterjee}, \citenamefont {Souma}, \citenamefont {Kaminski}, \citenamefont
  {Fretwell}, \citenamefont {Rosenkranz}, \citenamefont {Shi}, \citenamefont
  {Sato}, \citenamefont {Takahashi}, \citenamefont {Li}, \citenamefont {Raffy},
  \citenamefont {Kadowaki}, \citenamefont {Hinks}, \citenamefont {Ozyuzer},\
  and\ \citenamefont {Campuzano}}]{KanigelNatPhys2006}%
  \BibitemOpen
  \bibfield  {author} {\bibinfo {author} {\bibfnamefont {A.}~\bibnamefont
  {Kanigel}}, \bibinfo {author} {\bibfnamefont {M.~R.}\ \bibnamefont {Norman}},
  \bibinfo {author} {\bibfnamefont {M.}~\bibnamefont {Randeria}}, \bibinfo
  {author} {\bibfnamefont {U.}~\bibnamefont {Chatterjee}}, \bibinfo {author}
  {\bibfnamefont {S.}~\bibnamefont {Souma}}, \bibinfo {author} {\bibfnamefont
  {A.}~\bibnamefont {Kaminski}}, \bibinfo {author} {\bibfnamefont {H.~M.}\
  \bibnamefont {Fretwell}}, \bibinfo {author} {\bibfnamefont {S.}~\bibnamefont
  {Rosenkranz}}, \bibinfo {author} {\bibfnamefont {M.}~\bibnamefont {Shi}},
  \bibinfo {author} {\bibfnamefont {T.}~\bibnamefont {Sato}}, \bibinfo {author}
  {\bibfnamefont {T.}~\bibnamefont {Takahashi}}, \bibinfo {author}
  {\bibfnamefont {Z.~Z.}\ \bibnamefont {Li}}, \bibinfo {author} {\bibfnamefont
  {H.}~\bibnamefont {Raffy}}, \bibinfo {author} {\bibfnamefont
  {K.}~\bibnamefont {Kadowaki}}, \bibinfo {author} {\bibfnamefont
  {D.}~\bibnamefont {Hinks}}, \bibinfo {author} {\bibfnamefont
  {L.}~\bibnamefont {Ozyuzer}}, \ and\ \bibinfo {author} {\bibfnamefont
  {J.~C.}\ \bibnamefont {Campuzano}},\ }\bibfield  {title} {\enquote {\bibinfo
  {title} {Evolution of the pseudogap from {Fermi} arcs to the nodal liquid},}\
  }\href {\doibase 10.1038/nphys334} {\bibfield  {journal} {\bibinfo  {journal}
  {Nat. Phys.}\ }\textbf {\bibinfo {volume} {2}},\ \bibinfo {pages} {447--451}
  (\bibinfo {year} {2006})}\BibitemShut {NoStop}%
\bibitem [{\citenamefont {Kanigel}\ \emph {et~al.}(2007)\citenamefont
  {Kanigel}, \citenamefont {Chatterjee}, \citenamefont {Randeria},
  \citenamefont {Norman}, \citenamefont {Souma}, \citenamefont {Shi},
  \citenamefont {Li}, \citenamefont {Raffy},\ and\ \citenamefont
  {Campuzano}}]{KanigelPRL2007}%
  \BibitemOpen
  \bibfield  {author} {\bibinfo {author} {\bibfnamefont {A.}~\bibnamefont
  {Kanigel}}, \bibinfo {author} {\bibfnamefont {U.}~\bibnamefont {Chatterjee}},
  \bibinfo {author} {\bibfnamefont {M.}~\bibnamefont {Randeria}}, \bibinfo
  {author} {\bibfnamefont {M.~R.}\ \bibnamefont {Norman}}, \bibinfo {author}
  {\bibfnamefont {S.}~\bibnamefont {Souma}}, \bibinfo {author} {\bibfnamefont
  {M.}~\bibnamefont {Shi}}, \bibinfo {author} {\bibfnamefont {Z.~Z.}\
  \bibnamefont {Li}}, \bibinfo {author} {\bibfnamefont {H.}~\bibnamefont
  {Raffy}}, \ and\ \bibinfo {author} {\bibfnamefont {J.~C.}\ \bibnamefont
  {Campuzano}},\ }\bibfield  {title} {\enquote {\bibinfo {title} {Protected
  nodes and the collapse of {Fermi} arcs in high-{$T_c$} cuprate
  superconductors},}\ }\href {\doibase 10.1103/PhysRevLett.99.157001}
  {\bibfield  {journal} {\bibinfo  {journal} {Phys. Rev. Lett.}\ }\textbf
  {\bibinfo {volume} {99}},\ \bibinfo {pages} {157001} (\bibinfo {year}
  {2007})}\BibitemShut {NoStop}%
\bibitem [{Note1()}]{Note1}%
  \BibitemOpen
  \bibinfo {note} {The word pseudogap has been variously used to describe both
  spectral gap and Fermi arc phenomena; here we note the distinction between
  these phenomena.}\BibitemShut {Stop}%
\bibitem [{\citenamefont {Plat\'e}\ \emph {et~al.}(2005)\citenamefont
  {Plat\'e}, \citenamefont {Mottershead}, \citenamefont {Elfimov},
  \citenamefont {Peets}, \citenamefont {Liang}, \citenamefont {Bonn},
  \citenamefont {Hardy}, \citenamefont {Chiuzbaian}, \citenamefont {Falub},
  \citenamefont {Shi}, \citenamefont {Patthey},\ and\ \citenamefont
  {Damascelli}}]{PlatePRL2005}%
  \BibitemOpen
  \bibfield  {author} {\bibinfo {author} {\bibfnamefont {M.}~\bibnamefont
  {Plat\'e}}, \bibinfo {author} {\bibfnamefont {J.~D.~F.}\ \bibnamefont
  {Mottershead}}, \bibinfo {author} {\bibfnamefont {I.~S.}\ \bibnamefont
  {Elfimov}}, \bibinfo {author} {\bibfnamefont {D.~C.}\ \bibnamefont {Peets}},
  \bibinfo {author} {\bibfnamefont {R.}~\bibnamefont {Liang}}, \bibinfo
  {author} {\bibfnamefont {D.~A.}\ \bibnamefont {Bonn}}, \bibinfo {author}
  {\bibfnamefont {W.~N.}\ \bibnamefont {Hardy}}, \bibinfo {author}
  {\bibfnamefont {S.}~\bibnamefont {Chiuzbaian}}, \bibinfo {author}
  {\bibfnamefont {M.}~\bibnamefont {Falub}}, \bibinfo {author} {\bibfnamefont
  {M.}~\bibnamefont {Shi}}, \bibinfo {author} {\bibfnamefont {L.}~\bibnamefont
  {Patthey}}, \ and\ \bibinfo {author} {\bibfnamefont {A.}~\bibnamefont
  {Damascelli}},\ }\bibfield  {title} {\enquote {\bibinfo {title} {{Fermi}
  surface and quasiparticle excitations of overdoped
  {Tl$_2$Ba$_2$CuO$_{6+\delta}$}},}\ }\href {\doibase
  10.1103/PhysRevLett.95.077001} {\bibfield  {journal} {\bibinfo  {journal}
  {Phys. Rev. Lett.}\ }\textbf {\bibinfo {volume} {95}},\ \bibinfo {pages}
  {077001} (\bibinfo {year} {2005})}\BibitemShut {NoStop}%
\bibitem [{\citenamefont {Vignolle}\ \emph {et~al.}(2008)\citenamefont
  {Vignolle}, \citenamefont {Carrington}, \citenamefont {Cooper}, \citenamefont
  {French}, \citenamefont {Mackenzie}, \citenamefont {Jaudet}, \citenamefont
  {Vignolles}, \citenamefont {Proust},\ and\ \citenamefont
  {Hussey}}]{ProustNature2008}%
  \BibitemOpen
  \bibfield  {author} {\bibinfo {author} {\bibfnamefont {B.}~\bibnamefont
  {Vignolle}}, \bibinfo {author} {\bibfnamefont {A.}~\bibnamefont
  {Carrington}}, \bibinfo {author} {\bibfnamefont {R.~A.}\ \bibnamefont
  {Cooper}}, \bibinfo {author} {\bibfnamefont {M.~M.~J.}\ \bibnamefont
  {French}}, \bibinfo {author} {\bibfnamefont {A.~P.}\ \bibnamefont
  {Mackenzie}}, \bibinfo {author} {\bibfnamefont {C.}~\bibnamefont {Jaudet}},
  \bibinfo {author} {\bibfnamefont {D.}~\bibnamefont {Vignolles}}, \bibinfo
  {author} {\bibfnamefont {C.}~\bibnamefont {Proust}}, \ and\ \bibinfo {author}
  {\bibfnamefont {N.~E.}\ \bibnamefont {Hussey}},\ }\bibfield  {title}
  {\enquote {\bibinfo {title} {Quantum oscillations in an overdoped
  high-{$T_c$} superconductor},}\ }\href {\doibase 10.1038/nature07323}
  {\bibfield  {journal} {\bibinfo  {journal} {Nature}\ }\textbf {\bibinfo
  {volume} {455}},\ \bibinfo {pages} {952} (\bibinfo {year}
  {2008})}\BibitemShut {NoStop}%
\bibitem [{\citenamefont {Badoux}\ \emph
  {et~al.}(2016{\natexlab{a}})\citenamefont {Badoux}, \citenamefont {Tabis},
  \citenamefont {Lalibert{\'{e}}}, \citenamefont {Grissonnanche}, \citenamefont
  {Vignolle}, \citenamefont {Vignolles}, \citenamefont {B{\'{e}}ard},
  \citenamefont {Bonn}, \citenamefont {Hardy}, \citenamefont {Liang},
  \citenamefont {Doiron-Leyraud}, \citenamefont {Taillefer},\ and\
  \citenamefont {Proust}}]{BadouxNature2016}%
  \BibitemOpen
  \bibfield  {author} {\bibinfo {author} {\bibfnamefont {S.}~\bibnamefont
  {Badoux}}, \bibinfo {author} {\bibfnamefont {W.}~\bibnamefont {Tabis}},
  \bibinfo {author} {\bibfnamefont {F.}~\bibnamefont {Lalibert{\'{e}}}},
  \bibinfo {author} {\bibfnamefont {G.}~\bibnamefont {Grissonnanche}}, \bibinfo
  {author} {\bibfnamefont {B.}~\bibnamefont {Vignolle}}, \bibinfo {author}
  {\bibfnamefont {D.}~\bibnamefont {Vignolles}}, \bibinfo {author}
  {\bibfnamefont {J.}~\bibnamefont {B{\'{e}}ard}}, \bibinfo {author}
  {\bibfnamefont {D.~A.}\ \bibnamefont {Bonn}}, \bibinfo {author}
  {\bibfnamefont {W.~N.}\ \bibnamefont {Hardy}}, \bibinfo {author}
  {\bibfnamefont {R.}~\bibnamefont {Liang}}, \bibinfo {author} {\bibfnamefont
  {N.}~\bibnamefont {Doiron-Leyraud}}, \bibinfo {author} {\bibfnamefont
  {L.}~\bibnamefont {Taillefer}}, \ and\ \bibinfo {author} {\bibfnamefont
  {C.}~\bibnamefont {Proust}},\ }\bibfield  {title} {\enquote {\bibinfo {title}
  {{Change of carrier density at the pseudogap critical point of a cuprate
  superconductor}},}\ }\href {\doibase 10.1038/nature16983} {\bibfield
  {journal} {\bibinfo  {journal} {Nature}\ }\textbf {\bibinfo {volume} {531}},\
  \bibinfo {pages} {210--214} (\bibinfo {year}
  {2016}{\natexlab{a}})}\BibitemShut {NoStop}%
\bibitem [{\citenamefont {Collignon}\ \emph {et~al.}(2017)\citenamefont
  {Collignon}, \citenamefont {Badoux}, \citenamefont {Afshar}, \citenamefont
  {Michon}, \citenamefont {Lalibert{\'{e}}}, \citenamefont
  {Cyr-Choini{\`{e}}re}, \citenamefont {Zhou}, \citenamefont {Licciardello},
  \citenamefont {Wiedmann}, \citenamefont {Doiron-Leyraud},\ and\ \citenamefont
  {Taillefer}}]{CollignonPRB2017}%
  \BibitemOpen
  \bibfield  {author} {\bibinfo {author} {\bibfnamefont {C.}~\bibnamefont
  {Collignon}}, \bibinfo {author} {\bibfnamefont {S.}~\bibnamefont {Badoux}},
  \bibinfo {author} {\bibfnamefont {S.~A.~A.}\ \bibnamefont {Afshar}}, \bibinfo
  {author} {\bibfnamefont {B.}~\bibnamefont {Michon}}, \bibinfo {author}
  {\bibfnamefont {F.}~\bibnamefont {Lalibert{\'{e}}}}, \bibinfo {author}
  {\bibfnamefont {O.}~\bibnamefont {Cyr-Choini{\`{e}}re}}, \bibinfo {author}
  {\bibfnamefont {J.-S.}\ \bibnamefont {Zhou}}, \bibinfo {author}
  {\bibfnamefont {S.}~\bibnamefont {Licciardello}}, \bibinfo {author}
  {\bibfnamefont {S.}~\bibnamefont {Wiedmann}}, \bibinfo {author}
  {\bibfnamefont {N.}~\bibnamefont {Doiron-Leyraud}}, \ and\ \bibinfo {author}
  {\bibfnamefont {L.}~\bibnamefont {Taillefer}},\ }\bibfield  {title} {\enquote
  {\bibinfo {title} {{Fermi}-surface transformation across the pseudogap
  critical point of the cuprate superconductor
  {La$_{1.6-x}$Nd$_{0.4}$Sr$_x$CuO$_4$}},}\ }\href {\doibase
  10.1103/PhysRevB.95.224517} {\bibfield  {journal} {\bibinfo  {journal} {Phys.
  Rev. B}\ }\textbf {\bibinfo {volume} {95}},\ \bibinfo {pages} {224517}
  (\bibinfo {year} {2017})}\BibitemShut {NoStop}%
\bibitem [{\citenamefont {Legros}\ \emph {et~al.}(2019)\citenamefont {Legros},
  \citenamefont {Benhabib}, \citenamefont {Tabis}, \citenamefont
  {Lalibert{\'{e}}}, \citenamefont {Dion}, \citenamefont {Lizaire},
  \citenamefont {Vignolle}, \citenamefont {Vignolles}, \citenamefont {Raffy},
  \citenamefont {Li}, \citenamefont {Auban-Senzier}, \citenamefont
  {Doiron-Leyraud}, \citenamefont {Fournier}, \citenamefont {Colson},
  \citenamefont {Taillefer},\ and\ \citenamefont {Proust}}]{LegrosArXiv2018}%
  \BibitemOpen
  \bibfield  {author} {\bibinfo {author} {\bibfnamefont {A.}~\bibnamefont
  {Legros}}, \bibinfo {author} {\bibfnamefont {S.}~\bibnamefont {Benhabib}},
  \bibinfo {author} {\bibfnamefont {W.}~\bibnamefont {Tabis}}, \bibinfo
  {author} {\bibfnamefont {F.}~\bibnamefont {Lalibert{\'{e}}}}, \bibinfo
  {author} {\bibfnamefont {M.}~\bibnamefont {Dion}}, \bibinfo {author}
  {\bibfnamefont {M.}~\bibnamefont {Lizaire}}, \bibinfo {author} {\bibfnamefont
  {B.}~\bibnamefont {Vignolle}}, \bibinfo {author} {\bibfnamefont
  {D.}~\bibnamefont {Vignolles}}, \bibinfo {author} {\bibfnamefont
  {H.}~\bibnamefont {Raffy}}, \bibinfo {author} {\bibfnamefont {Z.~Z.}\
  \bibnamefont {Li}}, \bibinfo {author} {\bibfnamefont {P.}~\bibnamefont
  {Auban-Senzier}}, \bibinfo {author} {\bibfnamefont {N.}~\bibnamefont
  {Doiron-Leyraud}}, \bibinfo {author} {\bibfnamefont {P.}~\bibnamefont
  {Fournier}}, \bibinfo {author} {\bibfnamefont {D.}~\bibnamefont {Colson}},
  \bibinfo {author} {\bibfnamefont {L.}~\bibnamefont {Taillefer}}, \ and\
  \bibinfo {author} {\bibfnamefont {C.}~\bibnamefont {Proust}},\ }\bibfield
  {title} {\enquote {\bibinfo {title} {Universal {T}-linear resistivity and
  {Planckian} dissipation in overdoped cuprates},}\ }\href {\doibase
  10.1038/s41567-018-0334-2} {\bibfield  {journal} {\bibinfo  {journal} {Nat.
  Phys.}\ }\textbf {\bibinfo {volume} {15}},\ \bibinfo {pages} {142--147}
  (\bibinfo {year} {2019})}\BibitemShut {NoStop}%
\bibitem [{\citenamefont {Cooper}\ \emph {et~al.}(2009)\citenamefont {Cooper},
  \citenamefont {Wang}, \citenamefont {Vignolle}, \citenamefont {Lipscombe},
  \citenamefont {Hayden}, \citenamefont {Tanabe}, \citenamefont {Adachi},
  \citenamefont {Koike}, \citenamefont {Nohara}, \citenamefont {Takagi},
  \citenamefont {Proust},\ and\ \citenamefont {Hussey}}]{CooperScience2009}%
  \BibitemOpen
  \bibfield  {author} {\bibinfo {author} {\bibfnamefont {R.~A.}\ \bibnamefont
  {Cooper}}, \bibinfo {author} {\bibfnamefont {Y.}~\bibnamefont {Wang}},
  \bibinfo {author} {\bibfnamefont {B.}~\bibnamefont {Vignolle}}, \bibinfo
  {author} {\bibfnamefont {O.~J.}\ \bibnamefont {Lipscombe}}, \bibinfo {author}
  {\bibfnamefont {S.~M.}\ \bibnamefont {Hayden}}, \bibinfo {author}
  {\bibfnamefont {Y.}~\bibnamefont {Tanabe}}, \bibinfo {author} {\bibfnamefont
  {T.}~\bibnamefont {Adachi}}, \bibinfo {author} {\bibfnamefont
  {Y.}~\bibnamefont {Koike}}, \bibinfo {author} {\bibfnamefont
  {M.}~\bibnamefont {Nohara}}, \bibinfo {author} {\bibfnamefont
  {H.}~\bibnamefont {Takagi}}, \bibinfo {author} {\bibfnamefont
  {C.}~\bibnamefont {Proust}}, \ and\ \bibinfo {author} {\bibfnamefont {N.~E.}\
  \bibnamefont {Hussey}},\ }\bibfield  {title} {\enquote {\bibinfo {title}
  {Anomalous criticality in the electrical resistivity of
  {La$_{2-x}$Sr$_x$CuO$_4$}},}\ }\href {\doibase 10.1126/science.1165015}
  {\bibfield  {journal} {\bibinfo  {journal} {Science}\ }\textbf {\bibinfo
  {volume} {323}},\ \bibinfo {pages} {603--607} (\bibinfo {year}
  {2009})}\BibitemShut {NoStop}%
\bibitem [{\citenamefont {Bozovic}\ \emph {et~al.}(2016)\citenamefont
  {Bozovic}, \citenamefont {He}, \citenamefont {Wu},\ and\ \citenamefont
  {Bollinger}}]{BozovicNature2016}%
  \BibitemOpen
  \bibfield  {author} {\bibinfo {author} {\bibfnamefont {I.}~\bibnamefont
  {Bozovic}}, \bibinfo {author} {\bibfnamefont {X.}~\bibnamefont {He}},
  \bibinfo {author} {\bibfnamefont {J.}~\bibnamefont {Wu}}, \ and\ \bibinfo
  {author} {\bibfnamefont {A.~T.}\ \bibnamefont {Bollinger}},\ }\bibfield
  {title} {\enquote {\bibinfo {title} {Dependence of the critical temperature
  in overdoped copper oxides on superfluid density},}\ }\href {\doibase
  10.1038/nature19061} {\bibfield  {journal} {\bibinfo  {journal} {Nature}\
  }\textbf {\bibinfo {volume} {536}},\ \bibinfo {pages} {309} (\bibinfo {year}
  {2016})}\BibitemShut {NoStop}%
\bibitem [{\citenamefont {Wu}\ \emph {et~al.}(2017)\citenamefont {Wu},
  \citenamefont {Bollinger}, \citenamefont {He},\ and\ \citenamefont
  {Bozovic}}]{BozovicNature2017}%
  \BibitemOpen
  \bibfield  {author} {\bibinfo {author} {\bibfnamefont {J.}~\bibnamefont
  {Wu}}, \bibinfo {author} {\bibfnamefont {A.~T.}\ \bibnamefont {Bollinger}},
  \bibinfo {author} {\bibfnamefont {X.}~\bibnamefont {He}}, \ and\ \bibinfo
  {author} {\bibfnamefont {I.}~\bibnamefont {Bozovic}},\ }\bibfield  {title}
  {\enquote {\bibinfo {title} {Spontaneous breaking of rotational symmetry in
  copper oxide superconductors},}\ }\href {\doibase 10.1038/nature23290}
  {\bibfield  {journal} {\bibinfo  {journal} {Nature}\ }\textbf {\bibinfo
  {volume} {547}},\ \bibinfo {pages} {432--435} (\bibinfo {year}
  {2017})}\BibitemShut {NoStop}%
\bibitem [{\citenamefont {{Mahmood}}\ \emph {et~al.}(2019)\citenamefont
  {{Mahmood}}, \citenamefont {{He}}, \citenamefont {{Bo\u{z}ovi\'{c}}},\ and\
  \citenamefont {{Armitage}}}]{ArmitageArXiv2018}%
  \BibitemOpen
  \bibfield  {author} {\bibinfo {author} {\bibfnamefont {F.}~\bibnamefont
  {{Mahmood}}}, \bibinfo {author} {\bibfnamefont {X.}~\bibnamefont {{He}}},
  \bibinfo {author} {\bibfnamefont {I.}~\bibnamefont {{Bo\u{z}ovi\'{c}}}}, \
  and\ \bibinfo {author} {\bibfnamefont {N.~P.}\ \bibnamefont {{Armitage}}},\
  }\bibfield  {title} {\enquote {\bibinfo {title} {Locating the missing
  superconducting electrons in the overdoped cuprates
  {La$_{2-x}$Sr$_x$CuO$_4$}},}\ }\href {\doibase
  10.1103/PhysRevLett.122.027003} {\bibfield  {journal} {\bibinfo  {journal}
  {Phys. Rev. Lett.}\ }\textbf {\bibinfo {volume} {122}},\ \bibinfo {pages}
  {027003} (\bibinfo {year} {2019})}\BibitemShut {NoStop}%
\bibitem [{\citenamefont {Lee-Hone}\ \emph {et~al.}(2017)\citenamefont
  {Lee-Hone}, \citenamefont {Dodge},\ and\ \citenamefont
  {Broun}}]{Lee-HonePRB2017}%
  \BibitemOpen
  \bibfield  {author} {\bibinfo {author} {\bibfnamefont {N.~R.}\ \bibnamefont
  {Lee-Hone}}, \bibinfo {author} {\bibfnamefont {J.~S.}\ \bibnamefont {Dodge}},
  \ and\ \bibinfo {author} {\bibfnamefont {D.~M.}\ \bibnamefont {Broun}},\
  }\bibfield  {title} {\enquote {\bibinfo {title} {Disorder and superfluid
  density in overdoped cuprate superconductors},}\ }\href {\doibase
  10.1103/PhysRevB.96.024501} {\bibfield  {journal} {\bibinfo  {journal} {Phys.
  Rev. B}\ }\textbf {\bibinfo {volume} {96}},\ \bibinfo {pages} {024501}
  (\bibinfo {year} {2017})}\BibitemShut {NoStop}%
\bibitem [{\citenamefont {Lee-Hone}\ \emph {et~al.}(2018)\citenamefont
  {Lee-Hone}, \citenamefont {Mishra}, \citenamefont {Broun},\ and\
  \citenamefont {Hirschfeld}}]{Lee-HoneArXiv2018}%
  \BibitemOpen
  \bibfield  {author} {\bibinfo {author} {\bibfnamefont {N.~R.}\ \bibnamefont
  {Lee-Hone}}, \bibinfo {author} {\bibfnamefont {V.}~\bibnamefont {Mishra}},
  \bibinfo {author} {\bibfnamefont {D.~M.}\ \bibnamefont {Broun}}, \ and\
  \bibinfo {author} {\bibfnamefont {P.~J.}\ \bibnamefont {Hirschfeld}},\
  }\bibfield  {title} {\enquote {\bibinfo {title} {Optical conductivity of
  overdoped cuprate superconductors: {Application} to
  {La$_{2-x}$Sr$_x$CuO$_4$}},}\ }\href {\doibase 10.1103/PhysRevB.98.054506}
  {\bibfield  {journal} {\bibinfo  {journal} {Phys. Rev. B}\ }\textbf {\bibinfo
  {volume} {98}},\ \bibinfo {pages} {054506} (\bibinfo {year}
  {2018})}\BibitemShut {NoStop}%
\bibitem [{\citenamefont {{Le Tacon}}\ \emph {et~al.}(2013)\citenamefont {{Le
  Tacon}}, \citenamefont {Minola}, \citenamefont {Peets}, \citenamefont
  {{Moretti Sala}}, \citenamefont {{Blanco-Canosa}}, \citenamefont {Hinkov},
  \citenamefont {Liang}, \citenamefont {Bonn}, \citenamefont {Hardy},
  \citenamefont {Lin}, \citenamefont {Schmitt}, \citenamefont {Braicovich},
  \citenamefont {Ghiringhelli},\ and\ \citenamefont {Keimer}}]{LeTaconPRB2013}%
  \BibitemOpen
  \bibfield  {author} {\bibinfo {author} {\bibfnamefont {M.}~\bibnamefont {{Le
  Tacon}}}, \bibinfo {author} {\bibfnamefont {M.}~\bibnamefont {Minola}},
  \bibinfo {author} {\bibfnamefont {D.~C.}\ \bibnamefont {Peets}}, \bibinfo
  {author} {\bibfnamefont {M.}~\bibnamefont {{Moretti Sala}}}, \bibinfo
  {author} {\bibfnamefont {S.}~\bibnamefont {{Blanco-Canosa}}}, \bibinfo
  {author} {\bibfnamefont {V.}~\bibnamefont {Hinkov}}, \bibinfo {author}
  {\bibfnamefont {R.}~\bibnamefont {Liang}}, \bibinfo {author} {\bibfnamefont
  {D.~A.}\ \bibnamefont {Bonn}}, \bibinfo {author} {\bibfnamefont {W.~N.}\
  \bibnamefont {Hardy}}, \bibinfo {author} {\bibfnamefont {C.~T.}\ \bibnamefont
  {Lin}}, \bibinfo {author} {\bibfnamefont {T.}~\bibnamefont {Schmitt}},
  \bibinfo {author} {\bibfnamefont {L.}~\bibnamefont {Braicovich}}, \bibinfo
  {author} {\bibfnamefont {G.}~\bibnamefont {Ghiringhelli}}, \ and\ \bibinfo
  {author} {\bibfnamefont {B.}~\bibnamefont {Keimer}},\ }\bibfield  {title}
  {\enquote {\bibinfo {title} {Dispersive spin excitations in highly overdoped
  cuprates revealed by resonant inelastic x-ray scattering},}\ }\href {\doibase
  10.1103/PhysRevB.88.020501} {\bibfield  {journal} {\bibinfo  {journal} {Phys.
  Rev. B}\ }\textbf {\bibinfo {volume} {88}},\ \bibinfo {pages} {020501(R)}
  (\bibinfo {year} {2013})}\BibitemShut {NoStop}%
\bibitem [{\citenamefont {Dean}\ \emph {et~al.}(2013)\citenamefont {Dean},
  \citenamefont {Dellea}, \citenamefont {Springell}, \citenamefont
  {Yakhou-Harris}, \citenamefont {Kummer}, \citenamefont {Brookes},
  \citenamefont {Liu}, \citenamefont {Sun}, \citenamefont {Strle},
  \citenamefont {Schmitt}, \citenamefont {Braicovich}, \citenamefont
  {Ghiringhelli}, \citenamefont {Bo{\v{z}}ovi{\'{c}}},\ and\ \citenamefont
  {Hill}}]{DeanNatMater2013}%
  \BibitemOpen
  \bibfield  {author} {\bibinfo {author} {\bibfnamefont {M.~P.~M.}\
  \bibnamefont {Dean}}, \bibinfo {author} {\bibfnamefont {G.}~\bibnamefont
  {Dellea}}, \bibinfo {author} {\bibfnamefont {R.~S.}\ \bibnamefont
  {Springell}}, \bibinfo {author} {\bibfnamefont {F.}~\bibnamefont
  {Yakhou-Harris}}, \bibinfo {author} {\bibfnamefont {K.}~\bibnamefont
  {Kummer}}, \bibinfo {author} {\bibfnamefont {N.~B.}\ \bibnamefont {Brookes}},
  \bibinfo {author} {\bibfnamefont {X.}~\bibnamefont {Liu}}, \bibinfo {author}
  {\bibfnamefont {Y.~J.}\ \bibnamefont {Sun}}, \bibinfo {author} {\bibfnamefont
  {J.}~\bibnamefont {Strle}}, \bibinfo {author} {\bibfnamefont
  {T.}~\bibnamefont {Schmitt}}, \bibinfo {author} {\bibfnamefont
  {L.}~\bibnamefont {Braicovich}}, \bibinfo {author} {\bibfnamefont
  {G.}~\bibnamefont {Ghiringhelli}}, \bibinfo {author} {\bibfnamefont
  {I.}~\bibnamefont {Bo{\v{z}}ovi{\'{c}}}}, \ and\ \bibinfo {author}
  {\bibfnamefont {J.~P.}\ \bibnamefont {Hill}},\ }\bibfield  {title} {\enquote
  {\bibinfo {title} {Persistence of magnetic excitations in
  {La$_{2-x}$Sr$_x$CuO$_4$} from the undoped insulator to the heavily overdoped
  non-superconducting metal},}\ }\href {\doibase 10.1038/nmat3723} {\bibfield
  {journal} {\bibinfo  {journal} {Nat. Mater.}\ }\textbf {\bibinfo {volume}
  {12}},\ \bibinfo {pages} {1019--1023} (\bibinfo {year} {2013})}\BibitemShut
  {NoStop}%
\bibitem [{\citenamefont {Mross}\ and\ \citenamefont
  {Senthil}(2012{\natexlab{a}})}]{MrossPRL2012}%
  \BibitemOpen
  \bibfield  {author} {\bibinfo {author} {\bibfnamefont {D.~F.}\ \bibnamefont
  {Mross}}\ and\ \bibinfo {author} {\bibfnamefont {T.}~\bibnamefont
  {Senthil}},\ }\bibfield  {title} {\enquote {\bibinfo {title} {Theory of a
  continuous stripe melting transition in a two-dimensional metal: {A} possible
  application to cuprate superconductors},}\ }\href {\doibase
  10.1103/PhysRevLett.108.267001} {\bibfield  {journal} {\bibinfo  {journal}
  {Phys. Rev. Lett.}\ }\textbf {\bibinfo {volume} {108}},\ \bibinfo {pages}
  {267001} (\bibinfo {year} {2012}{\natexlab{a}})}\BibitemShut {NoStop}%
\bibitem [{\citenamefont {Mross}\ and\ \citenamefont
  {Senthil}(2012{\natexlab{b}})}]{MrossPRB2012}%
  \BibitemOpen
  \bibfield  {author} {\bibinfo {author} {\bibfnamefont {D.~F.}\ \bibnamefont
  {Mross}}\ and\ \bibinfo {author} {\bibfnamefont {T.}~\bibnamefont
  {Senthil}},\ }\bibfield  {title} {\enquote {\bibinfo {title} {Stripe melting
  and quantum criticality in correlated metals},}\ }\href {\doibase
  10.1103/PhysRevB.86.115138} {\bibfield  {journal} {\bibinfo  {journal} {Phys.
  Rev. B}\ }\textbf {\bibinfo {volume} {86}},\ \bibinfo {pages} {115138}
  (\bibinfo {year} {2012}{\natexlab{b}})}\BibitemShut {NoStop}%
\bibitem [{\citenamefont {Fujita}\ \emph
  {et~al.}(2014{\natexlab{a}})\citenamefont {Fujita}, \citenamefont {Kim},
  \citenamefont {Lee}, \citenamefont {Lee}, \citenamefont {Hamidian},
  \citenamefont {Firmo}, \citenamefont {Mukhopadhyay}, \citenamefont {Eisaki},
  \citenamefont {Uchida}, \citenamefont {Lawler}, \citenamefont {Kim},\ and\
  \citenamefont {Davis}}]{FujitaScience2014}%
  \BibitemOpen
  \bibfield  {author} {\bibinfo {author} {\bibfnamefont {K.}~\bibnamefont
  {Fujita}}, \bibinfo {author} {\bibfnamefont {C.~K.}\ \bibnamefont {Kim}},
  \bibinfo {author} {\bibfnamefont {I.}~\bibnamefont {Lee}}, \bibinfo {author}
  {\bibfnamefont {J.}~\bibnamefont {Lee}}, \bibinfo {author} {\bibfnamefont
  {M.~H.}\ \bibnamefont {Hamidian}}, \bibinfo {author} {\bibfnamefont {I.~A.}\
  \bibnamefont {Firmo}}, \bibinfo {author} {\bibfnamefont {S.}~\bibnamefont
  {Mukhopadhyay}}, \bibinfo {author} {\bibfnamefont {H.}~\bibnamefont
  {Eisaki}}, \bibinfo {author} {\bibfnamefont {S.}~\bibnamefont {Uchida}},
  \bibinfo {author} {\bibfnamefont {M.~J.}\ \bibnamefont {Lawler}}, \bibinfo
  {author} {\bibfnamefont {E.-A.}\ \bibnamefont {Kim}}, \ and\ \bibinfo
  {author} {\bibfnamefont {J.C.}\ \bibnamefont {Davis}},\ }\bibfield  {title}
  {\enquote {\bibinfo {title} {Simultaneous transitions in cuprate
  momentum-space topology and electronic symmetry breaking},}\ }\href {\doibase
  10.1126/science.1248783} {\bibfield  {journal} {\bibinfo  {journal}
  {Science}\ }\textbf {\bibinfo {volume} {344}},\ \bibinfo {pages} {612--616}
  (\bibinfo {year} {2014}{\natexlab{a}})}\BibitemShut {NoStop}%
\bibitem [{\citenamefont {Badoux}\ \emph
  {et~al.}(2016{\natexlab{b}})\citenamefont {Badoux}, \citenamefont {Afshar},
  \citenamefont {Michon}, \citenamefont {Ouellet}, \citenamefont {Fortier},
  \citenamefont {LeBoeuf}, \citenamefont {Croft}, \citenamefont {Lester},
  \citenamefont {Hayden}, \citenamefont {Takagi}, \citenamefont {Yamada},
  \citenamefont {Graf}, \citenamefont {Doiron-Leyraud},\ and\ \citenamefont
  {Taillefer}}]{BadouxPRX2016}%
  \BibitemOpen
  \bibfield  {author} {\bibinfo {author} {\bibfnamefont {S.}~\bibnamefont
  {Badoux}}, \bibinfo {author} {\bibfnamefont {S.~A.~A.}\ \bibnamefont
  {Afshar}}, \bibinfo {author} {\bibfnamefont {B.}~\bibnamefont {Michon}},
  \bibinfo {author} {\bibfnamefont {A.}~\bibnamefont {Ouellet}}, \bibinfo
  {author} {\bibfnamefont {S.}~\bibnamefont {Fortier}}, \bibinfo {author}
  {\bibfnamefont {D.}~\bibnamefont {LeBoeuf}}, \bibinfo {author} {\bibfnamefont
  {T.~P.}\ \bibnamefont {Croft}}, \bibinfo {author} {\bibfnamefont
  {C.}~\bibnamefont {Lester}}, \bibinfo {author} {\bibfnamefont {S.~M.}\
  \bibnamefont {Hayden}}, \bibinfo {author} {\bibfnamefont {H.}~\bibnamefont
  {Takagi}}, \bibinfo {author} {\bibfnamefont {K.}~\bibnamefont {Yamada}},
  \bibinfo {author} {\bibfnamefont {D.}~\bibnamefont {Graf}}, \bibinfo {author}
  {\bibfnamefont {N.}~\bibnamefont {Doiron-Leyraud}}, \ and\ \bibinfo {author}
  {\bibfnamefont {L.}~\bibnamefont {Taillefer}},\ }\bibfield  {title} {\enquote
  {\bibinfo {title} {Critical doping for the onset of {Fermi}-surface
  reconstruction by charge-density-wave order in the cuprate superconductor
  {La$_{2-x}$Sr$_x$CuO$_4$}},}\ }\href {\doibase 10.1103/PhysRevX.6.021004}
  {\bibfield  {journal} {\bibinfo  {journal} {Phys. Rev. X}\ }\textbf {\bibinfo
  {volume} {6}},\ \bibinfo {pages} {021004} (\bibinfo {year}
  {2016}{\natexlab{b}})}\BibitemShut {NoStop}%
\bibitem [{\citenamefont {He}\ \emph {et~al.}(2014)\citenamefont {He},
  \citenamefont {Yin}, \citenamefont {Zech}, \citenamefont {Soumyanarayanan},
  \citenamefont {Yee}, \citenamefont {Williams}, \citenamefont {Boyer},
  \citenamefont {Chatterjee}, \citenamefont {Wise}, \citenamefont {Zeljkovic},
  \citenamefont {Kondo}, \citenamefont {Takeuchi}, \citenamefont {Ikuta},
  \citenamefont {Mistark}, \citenamefont {Markiewicz}, \citenamefont {Bansil},
  \citenamefont {Sachdev}, \citenamefont {Hudson},\ and\ \citenamefont
  {Hoffman}}]{HeScience2014}%
  \BibitemOpen
  \bibfield  {author} {\bibinfo {author} {\bibfnamefont {Y.}~\bibnamefont
  {He}}, \bibinfo {author} {\bibfnamefont {Y.}~\bibnamefont {Yin}}, \bibinfo
  {author} {\bibfnamefont {M.}~\bibnamefont {Zech}}, \bibinfo {author}
  {\bibfnamefont {A.}~\bibnamefont {Soumyanarayanan}}, \bibinfo {author}
  {\bibfnamefont {M.~M.}\ \bibnamefont {Yee}}, \bibinfo {author} {\bibfnamefont
  {T.}~\bibnamefont {Williams}}, \bibinfo {author} {\bibfnamefont {M.~C.}\
  \bibnamefont {Boyer}}, \bibinfo {author} {\bibfnamefont {K.}~\bibnamefont
  {Chatterjee}}, \bibinfo {author} {\bibfnamefont {W.~D.}\ \bibnamefont
  {Wise}}, \bibinfo {author} {\bibfnamefont {I.}~\bibnamefont {Zeljkovic}},
  \bibinfo {author} {\bibfnamefont {T.}~\bibnamefont {Kondo}}, \bibinfo
  {author} {\bibfnamefont {T.}~\bibnamefont {Takeuchi}}, \bibinfo {author}
  {\bibfnamefont {H.}~\bibnamefont {Ikuta}}, \bibinfo {author} {\bibfnamefont
  {P.}~\bibnamefont {Mistark}}, \bibinfo {author} {\bibfnamefont {R.~S.}\
  \bibnamefont {Markiewicz}}, \bibinfo {author} {\bibfnamefont
  {A.}~\bibnamefont {Bansil}}, \bibinfo {author} {\bibfnamefont
  {S.}~\bibnamefont {Sachdev}}, \bibinfo {author} {\bibfnamefont {E.~W.}\
  \bibnamefont {Hudson}}, \ and\ \bibinfo {author} {\bibfnamefont {J.~E.}\
  \bibnamefont {Hoffman}},\ }\bibfield  {title} {\enquote {\bibinfo {title}
  {{Fermi} surface and pseudogap evolution in a cuprate superconductor},}\
  }\href {\doibase 10.1126/science.1248221} {\bibfield  {journal} {\bibinfo
  {journal} {Science}\ }\textbf {\bibinfo {volume} {344}},\ \bibinfo {pages}
  {608--611} (\bibinfo {year} {2014})}\BibitemShut {NoStop}%
\bibitem [{\citenamefont {Peng}\ \emph {et~al.}(2018)\citenamefont {Peng},
  \citenamefont {Fumagalli}, \citenamefont {Ding}, \citenamefont {Minola},
  \citenamefont {Caprara}, \citenamefont {Betto}, \citenamefont {Bluschke},
  \citenamefont {{De Luca}}, \citenamefont {Kummer}, \citenamefont
  {Lefran{\c{c}}ois}, \citenamefont {Salluzzo}, \citenamefont {Suzuki},
  \citenamefont {{Le Tacon}}, \citenamefont {Zhou}, \citenamefont {Brookes},
  \citenamefont {Keimer}, \citenamefont {Braicovich}, \citenamefont {Grilli},\
  and\ \citenamefont {Ghiringhelli}}]{PengNatMater2017}%
  \BibitemOpen
  \bibfield  {author} {\bibinfo {author} {\bibfnamefont {Y.~Y.}\ \bibnamefont
  {Peng}}, \bibinfo {author} {\bibfnamefont {R.}~\bibnamefont {Fumagalli}},
  \bibinfo {author} {\bibfnamefont {Y.}~\bibnamefont {Ding}}, \bibinfo {author}
  {\bibfnamefont {M.}~\bibnamefont {Minola}}, \bibinfo {author} {\bibfnamefont
  {S.}~\bibnamefont {Caprara}}, \bibinfo {author} {\bibfnamefont
  {D.}~\bibnamefont {Betto}}, \bibinfo {author} {\bibfnamefont
  {M.}~\bibnamefont {Bluschke}}, \bibinfo {author} {\bibfnamefont {G.~M.}\
  \bibnamefont {{De Luca}}}, \bibinfo {author} {\bibfnamefont {K.}~\bibnamefont
  {Kummer}}, \bibinfo {author} {\bibfnamefont {E.}~\bibnamefont
  {Lefran{\c{c}}ois}}, \bibinfo {author} {\bibfnamefont {M.}~\bibnamefont
  {Salluzzo}}, \bibinfo {author} {\bibfnamefont {H.}~\bibnamefont {Suzuki}},
  \bibinfo {author} {\bibfnamefont {M.}~\bibnamefont {{Le Tacon}}}, \bibinfo
  {author} {\bibfnamefont {X.~J.}\ \bibnamefont {Zhou}}, \bibinfo {author}
  {\bibfnamefont {N.~B.}\ \bibnamefont {Brookes}}, \bibinfo {author}
  {\bibfnamefont {B.}~\bibnamefont {Keimer}}, \bibinfo {author} {\bibfnamefont
  {L.}~\bibnamefont {Braicovich}}, \bibinfo {author} {\bibfnamefont
  {M.}~\bibnamefont {Grilli}}, \ and\ \bibinfo {author} {\bibfnamefont
  {G.}~\bibnamefont {Ghiringhelli}},\ }\bibfield  {title} {\enquote {\bibinfo
  {title} {Re-entrant charge order in overdoped
  {(Bi,Pb)$_{2.12}$Sr$_{1.88}$CuO$_{6+\delta}$} outside the pseudogap
  regime},}\ }\href {\doibase 10.1038/s41563-018-0108-3} {\bibfield  {journal}
  {\bibinfo  {journal} {Nat. Mater.}\ }\textbf {\bibinfo {volume} {17}},\
  \bibinfo {pages} {697--702} (\bibinfo {year} {2018})}\BibitemShut {NoStop}%
\bibitem [{Note2()}]{Note2}%
  \BibitemOpen
  \bibinfo {note} {To the authors' knowledge, the FS transition in Bi2201, near
  $p=0.14$, has been reported only by scanning tunneling microscopy~\cite
  {HeScience2014}, but Kondo \protect \textit {et al.}'s ARPES measurements
  (compare Figs. 3e and f of Ref.\ \protect \rev@citealp {KondoNature2009})
  also show that the spectral weight in the sharp quasiparticle peak at the
  antinode vanishes between the optimally doped and underdoped UD23K
  compounds.}\BibitemShut {Stop}%
\bibitem [{\citenamefont {Kondo}\ \emph {et~al.}(2011)\citenamefont {Kondo},
  \citenamefont {Hamaya}, \citenamefont {Palczewski}, \citenamefont {Takeuchi},
  \citenamefont {Wen}, \citenamefont {Xu}, \citenamefont {Gu}, \citenamefont
  {Schmalian},\ and\ \citenamefont {Kaminski}}]{KondoNatPhys2011}%
  \BibitemOpen
  \bibfield  {author} {\bibinfo {author} {\bibfnamefont {T.}~\bibnamefont
  {Kondo}}, \bibinfo {author} {\bibfnamefont {Y.}~\bibnamefont {Hamaya}},
  \bibinfo {author} {\bibfnamefont {A.~D.}\ \bibnamefont {Palczewski}},
  \bibinfo {author} {\bibfnamefont {T.}~\bibnamefont {Takeuchi}}, \bibinfo
  {author} {\bibfnamefont {J.~S.}\ \bibnamefont {Wen}}, \bibinfo {author}
  {\bibfnamefont {Z.~J.}\ \bibnamefont {Xu}}, \bibinfo {author} {\bibfnamefont
  {G.}~\bibnamefont {Gu}}, \bibinfo {author} {\bibfnamefont {J.}~\bibnamefont
  {Schmalian}}, \ and\ \bibinfo {author} {\bibfnamefont {A.}~\bibnamefont
  {Kaminski}},\ }\bibfield  {title} {\enquote {\bibinfo {title} {Disentangling
  {Cooper}-pair formation above the transition temperature from the pseudogap
  state in the cuprates},}\ }\href {\doibase 10.1038/nphys1851} {\bibfield
  {journal} {\bibinfo  {journal} {Nat. Phys.}\ }\textbf {\bibinfo {volume}
  {7}},\ \bibinfo {pages} {21--25} (\bibinfo {year} {2011})}\BibitemShut
  {NoStop}%
\bibitem [{\citenamefont {Zheng}\ \emph {et~al.}(2005)\citenamefont {Zheng},
  \citenamefont {Kuhns}, \citenamefont {Reyes}, \citenamefont {Liang},\ and\
  \citenamefont {Lin}}]{ZhengPRL2005}%
  \BibitemOpen
  \bibfield  {author} {\bibinfo {author} {\bibfnamefont {G.-Q.}\ \bibnamefont
  {Zheng}}, \bibinfo {author} {\bibfnamefont {P.~L.}\ \bibnamefont {Kuhns}},
  \bibinfo {author} {\bibfnamefont {A.~P.}\ \bibnamefont {Reyes}}, \bibinfo
  {author} {\bibfnamefont {B.}~\bibnamefont {Liang}}, \ and\ \bibinfo {author}
  {\bibfnamefont {C.~T.}\ \bibnamefont {Lin}},\ }\bibfield  {title} {\enquote
  {\bibinfo {title} {Critical point and the nature of the pseudogap of
  single-layered copper-oxide {Bi$_2$Sr$_{2-x}$La$_x$CuO$_{6+\delta}$}},}\
  }\href {\doibase 10.1103/PhysRevLett.94.047006} {\bibfield  {journal}
  {\bibinfo  {journal} {Phys. Rev. Lett.}\ }\textbf {\bibinfo {volume} {94}},\
  \bibinfo {pages} {047006} (\bibinfo {year} {2005})}\BibitemShut {NoStop}%
\bibitem [{\citenamefont {Comin}\ \emph {et~al.}(2014)\citenamefont {Comin},
  \citenamefont {Frano}, \citenamefont {Yee}, \citenamefont {Yoshida},
  \citenamefont {Eisaki}, \citenamefont {Schierle}, \citenamefont {Weschke},
  \citenamefont {Sutarto}, \citenamefont {He}, \citenamefont {Soumyanarayanan},
  \citenamefont {He}, \citenamefont {{Le Tacon}}, \citenamefont {Elfimov},
  \citenamefont {Hoffman}, \citenamefont {Sawatzky}, \citenamefont {Keimer},\
  and\ \citenamefont {Damascelli}}]{CominScience2014}%
  \BibitemOpen
  \bibfield  {author} {\bibinfo {author} {\bibfnamefont {R.}~\bibnamefont
  {Comin}}, \bibinfo {author} {\bibfnamefont {A.}~\bibnamefont {Frano}},
  \bibinfo {author} {\bibfnamefont {M.~M.}\ \bibnamefont {Yee}}, \bibinfo
  {author} {\bibfnamefont {Y.}~\bibnamefont {Yoshida}}, \bibinfo {author}
  {\bibfnamefont {H.}~\bibnamefont {Eisaki}}, \bibinfo {author} {\bibfnamefont
  {E.}~\bibnamefont {Schierle}}, \bibinfo {author} {\bibfnamefont
  {E.}~\bibnamefont {Weschke}}, \bibinfo {author} {\bibfnamefont
  {R.}~\bibnamefont {Sutarto}}, \bibinfo {author} {\bibfnamefont
  {F.}~\bibnamefont {He}}, \bibinfo {author} {\bibfnamefont {A.}~\bibnamefont
  {Soumyanarayanan}}, \bibinfo {author} {\bibfnamefont {Y.}~\bibnamefont {He}},
  \bibinfo {author} {\bibfnamefont {M.}~\bibnamefont {{Le Tacon}}}, \bibinfo
  {author} {\bibfnamefont {I.~S.}\ \bibnamefont {Elfimov}}, \bibinfo {author}
  {\bibfnamefont {J.~E.}\ \bibnamefont {Hoffman}}, \bibinfo {author}
  {\bibfnamefont {G.~A.}\ \bibnamefont {Sawatzky}}, \bibinfo {author}
  {\bibfnamefont {B.}~\bibnamefont {Keimer}}, \ and\ \bibinfo {author}
  {\bibfnamefont {A.}~\bibnamefont {Damascelli}},\ }\bibfield  {title}
  {\enquote {\bibinfo {title} {Charge order driven by {Fermi}-arc instability
  in {Bi$_2$Sr$_{2-x}$La$_x$CuO$_{6+\delta}$}},}\ }\href {\doibase
  10.1126/science.1242996} {\bibfield  {journal} {\bibinfo  {journal}
  {Science}\ }\textbf {\bibinfo {volume} {343}},\ \bibinfo {pages} {390--392}
  (\bibinfo {year} {2014})}\BibitemShut {NoStop}%
\bibitem [{\citenamefont {Peng}\ \emph {et~al.}(2016)\citenamefont {Peng},
  \citenamefont {Salluzzo}, \citenamefont {Sun}, \citenamefont {Ponti},
  \citenamefont {Betto}, \citenamefont {Ferretti}, \citenamefont {Fumagalli},
  \citenamefont {Kummer}, \citenamefont {{Le Tacon}}, \citenamefont {Zhou},
  \citenamefont {Brookes}, \citenamefont {Braicovich},\ and\ \citenamefont
  {Ghiringhelli}}]{PengPRB2016}%
  \BibitemOpen
  \bibfield  {author} {\bibinfo {author} {\bibfnamefont {Y.~Y.}\ \bibnamefont
  {Peng}}, \bibinfo {author} {\bibfnamefont {M.}~\bibnamefont {Salluzzo}},
  \bibinfo {author} {\bibfnamefont {X.}~\bibnamefont {Sun}}, \bibinfo {author}
  {\bibfnamefont {A.}~\bibnamefont {Ponti}}, \bibinfo {author} {\bibfnamefont
  {D.}~\bibnamefont {Betto}}, \bibinfo {author} {\bibfnamefont {A.~M.}\
  \bibnamefont {Ferretti}}, \bibinfo {author} {\bibfnamefont {F.}~\bibnamefont
  {Fumagalli}}, \bibinfo {author} {\bibfnamefont {K.}~\bibnamefont {Kummer}},
  \bibinfo {author} {\bibfnamefont {M.}~\bibnamefont {{Le Tacon}}}, \bibinfo
  {author} {\bibfnamefont {X.~J.}\ \bibnamefont {Zhou}}, \bibinfo {author}
  {\bibfnamefont {N.~B.}\ \bibnamefont {Brookes}}, \bibinfo {author}
  {\bibfnamefont {L.}~\bibnamefont {Braicovich}}, \ and\ \bibinfo {author}
  {\bibfnamefont {G.}~\bibnamefont {Ghiringhelli}},\ }\bibfield  {title}
  {\enquote {\bibinfo {title} {Direct observation of charge order in underdoped
  and optimally doped {Bi$_2$(Sr,La)$_2$CuO$_{6+\delta}$} by resonant inelastic
  x-ray scattering},}\ }\href {\doibase 10.1103/PhysRevB.94.184511} {\bibfield
  {journal} {\bibinfo  {journal} {Phys. Rev. B}\ }\textbf {\bibinfo {volume}
  {94}},\ \bibinfo {pages} {184511} (\bibinfo {year} {2016})}\BibitemShut
  {NoStop}%
\bibitem [{\citenamefont {Wise}\ \emph {et~al.}(2008)\citenamefont {Wise},
  \citenamefont {Boyer}, \citenamefont {Chatterjee}, \citenamefont {Kondo},
  \citenamefont {Takeuchi}, \citenamefont {Ikuta}, \citenamefont {Wang},\ and\
  \citenamefont {Hudson}}]{WiseNatPhys2008}%
  \BibitemOpen
  \bibfield  {author} {\bibinfo {author} {\bibfnamefont {W.~D.}\ \bibnamefont
  {Wise}}, \bibinfo {author} {\bibfnamefont {M.~C.}\ \bibnamefont {Boyer}},
  \bibinfo {author} {\bibfnamefont {K.}~\bibnamefont {Chatterjee}}, \bibinfo
  {author} {\bibfnamefont {T.}~\bibnamefont {Kondo}}, \bibinfo {author}
  {\bibfnamefont {T.}~\bibnamefont {Takeuchi}}, \bibinfo {author}
  {\bibfnamefont {H.}~\bibnamefont {Ikuta}}, \bibinfo {author} {\bibfnamefont
  {Y.}~\bibnamefont {Wang}}, \ and\ \bibinfo {author} {\bibfnamefont {E.~W.}\
  \bibnamefont {Hudson}},\ }\bibfield  {title} {\enquote {\bibinfo {title}
  {Charge-density-wave origin of cuprate checkerboard visualized by scanning
  tunnelling microscopy},}\ }\href {\doibase 10.1038/nphys1021} {\bibfield
  {journal} {\bibinfo  {journal} {Nat. Phys.}\ }\textbf {\bibinfo {volume}
  {4}},\ \bibinfo {pages} {696--699} (\bibinfo {year} {2008})}\BibitemShut
  {NoStop}%
\bibitem [{\citenamefont {Wise}\ \emph {et~al.}(2009)\citenamefont {Wise},
  \citenamefont {Chatterjee}, \citenamefont {Boyer}, \citenamefont {Kondo},
  \citenamefont {Takeuchi}, \citenamefont {Ikuta}, \citenamefont {Xu},
  \citenamefont {Wen}, \citenamefont {Gu}, \citenamefont {Wang},\ and\
  \citenamefont {Hudson}}]{WiseNatPhys2009}%
  \BibitemOpen
  \bibfield  {author} {\bibinfo {author} {\bibfnamefont {W.~D.}\ \bibnamefont
  {Wise}}, \bibinfo {author} {\bibfnamefont {K.}~\bibnamefont {Chatterjee}},
  \bibinfo {author} {\bibfnamefont {M.~C.}\ \bibnamefont {Boyer}}, \bibinfo
  {author} {\bibfnamefont {T.}~\bibnamefont {Kondo}}, \bibinfo {author}
  {\bibfnamefont {T.}~\bibnamefont {Takeuchi}}, \bibinfo {author}
  {\bibfnamefont {H.}~\bibnamefont {Ikuta}}, \bibinfo {author} {\bibfnamefont
  {Z.}~\bibnamefont {Xu}}, \bibinfo {author} {\bibfnamefont {J.}~\bibnamefont
  {Wen}}, \bibinfo {author} {\bibfnamefont {G.~D.}\ \bibnamefont {Gu}},
  \bibinfo {author} {\bibfnamefont {Y.}~\bibnamefont {Wang}}, \ and\ \bibinfo
  {author} {\bibfnamefont {E.~W.}\ \bibnamefont {Hudson}},\ }\bibfield  {title}
  {\enquote {\bibinfo {title} {Imaging nanoscale {Fermi}-surface variations in
  an inhomogeneous superconductor},}\ }\href {\doibase 10.1038/nphys1197}
  {\bibfield  {journal} {\bibinfo  {journal} {Nat. Phys.}\ }\textbf {\bibinfo
  {volume} {5}},\ \bibinfo {pages} {213--216} (\bibinfo {year}
  {2009})}\BibitemShut {NoStop}%
\bibitem [{\citenamefont {Cai}\ \emph {et~al.}(2016)\citenamefont {Cai},
  \citenamefont {Ruan}, \citenamefont {Peng}, \citenamefont {Ye}, \citenamefont
  {Li}, \citenamefont {Hao}, \citenamefont {Zhou}, \citenamefont {Lee},\ and\
  \citenamefont {Wang}}]{CaiNatPhys2016}%
  \BibitemOpen
  \bibfield  {author} {\bibinfo {author} {\bibfnamefont {P.}~\bibnamefont
  {Cai}}, \bibinfo {author} {\bibfnamefont {W.}~\bibnamefont {Ruan}}, \bibinfo
  {author} {\bibfnamefont {Y.}~\bibnamefont {Peng}}, \bibinfo {author}
  {\bibfnamefont {C.}~\bibnamefont {Ye}}, \bibinfo {author} {\bibfnamefont
  {X.}~\bibnamefont {Li}}, \bibinfo {author} {\bibfnamefont {Z.}~\bibnamefont
  {Hao}}, \bibinfo {author} {\bibfnamefont {X.}~\bibnamefont {Zhou}}, \bibinfo
  {author} {\bibfnamefont {D.-H.}\ \bibnamefont {Lee}}, \ and\ \bibinfo
  {author} {\bibfnamefont {Y.}~\bibnamefont {Wang}},\ }\bibfield  {title}
  {\enquote {\bibinfo {title} {{Visualizing the evolution from the Mott
  insulator to a charge-ordered insulator in lightly doped cuprates}},}\ }\href
  {\doibase 10.1038/nphys3840} {\bibfield  {journal} {\bibinfo  {journal} {Nat.
  Phys.}\ }\textbf {\bibinfo {volume} {12}},\ \bibinfo {pages} {1047--1051}
  (\bibinfo {year} {2016})}\BibitemShut {NoStop}%
\bibitem [{\citenamefont {Ando}\ \emph {et~al.}(2000)\citenamefont {Ando},
  \citenamefont {Hanaki}, \citenamefont {Ono}, \citenamefont {Murayama},
  \citenamefont {Segawa}, \citenamefont {Miyamoto},\ and\ \citenamefont
  {Komiya}}]{AndoPRB2000}%
  \BibitemOpen
  \bibfield  {author} {\bibinfo {author} {\bibfnamefont {Y.}~\bibnamefont
  {Ando}}, \bibinfo {author} {\bibfnamefont {Y.}~\bibnamefont {Hanaki}},
  \bibinfo {author} {\bibfnamefont {S.}~\bibnamefont {Ono}}, \bibinfo {author}
  {\bibfnamefont {T.}~\bibnamefont {Murayama}}, \bibinfo {author}
  {\bibfnamefont {K.}~\bibnamefont {Segawa}}, \bibinfo {author} {\bibfnamefont
  {N.}~\bibnamefont {Miyamoto}}, \ and\ \bibinfo {author} {\bibfnamefont
  {S.}~\bibnamefont {Komiya}},\ }\bibfield  {title} {\enquote {\bibinfo {title}
  {Carrier concentrations in {Bi$_2$Sr$_{2-z}$La$_z$CuO$_{6+\delta}$} single
  crystals and their relation to the {Hall} coefficient and thermopower},}\
  }\href {\doibase 10.1103/PhysRevB.61.R14956} {\bibfield  {journal} {\bibinfo
  {journal} {Phys. Rev. B}\ }\textbf {\bibinfo {volume} {61}},\ \bibinfo
  {pages} {R14956} (\bibinfo {year} {2000})}\BibitemShut {NoStop}%
\bibitem [{\citenamefont {Zeljkovic}\ \emph {et~al.}(2012)\citenamefont
  {Zeljkovic}, \citenamefont {Xu}, \citenamefont {Wen}, \citenamefont {Gu},
  \citenamefont {Markiewicz},\ and\ \citenamefont
  {Hoffman}}]{ZeljkovicScience2012}%
  \BibitemOpen
  \bibfield  {author} {\bibinfo {author} {\bibfnamefont {I.}~\bibnamefont
  {Zeljkovic}}, \bibinfo {author} {\bibfnamefont {Z.}~\bibnamefont {Xu}},
  \bibinfo {author} {\bibfnamefont {J.}~\bibnamefont {Wen}}, \bibinfo {author}
  {\bibfnamefont {G.}~\bibnamefont {Gu}}, \bibinfo {author} {\bibfnamefont
  {R.~S.}\ \bibnamefont {Markiewicz}}, \ and\ \bibinfo {author} {\bibfnamefont
  {J.~E.}\ \bibnamefont {Hoffman}},\ }\bibfield  {title} {\enquote {\bibinfo
  {title} {Imaging the impact of single oxygen atoms on superconducting
  {Bi$_{2+y}$Sr$_{2-y}$CaCu$_2$O$_{8+x}$}},}\ }\href {\doibase
  10.1126/science.1218648} {\bibfield  {journal} {\bibinfo  {journal}
  {Science}\ }\textbf {\bibinfo {volume} {337}},\ \bibinfo {pages} {320--323}
  (\bibinfo {year} {2012})}\BibitemShut {NoStop}%
\bibitem [{\citenamefont {McElroy}\ \emph
  {et~al.}(2005{\natexlab{a}})\citenamefont {McElroy}, \citenamefont {Lee},
  \citenamefont {Hoffman}, \citenamefont {Lang}, \citenamefont {Lee},
  \citenamefont {Hudson}, \citenamefont {Eisaki}, \citenamefont {Uchida},\ and\
  \citenamefont {Davis}}]{McElroyPRL2005}%
  \BibitemOpen
  \bibfield  {author} {\bibinfo {author} {\bibfnamefont {K.}~\bibnamefont
  {McElroy}}, \bibinfo {author} {\bibfnamefont {D.-H.}\ \bibnamefont {Lee}},
  \bibinfo {author} {\bibfnamefont {J.~E.}\ \bibnamefont {Hoffman}}, \bibinfo
  {author} {\bibfnamefont {K.~M.}\ \bibnamefont {Lang}}, \bibinfo {author}
  {\bibfnamefont {J.}~\bibnamefont {Lee}}, \bibinfo {author} {\bibfnamefont
  {E.~W.}\ \bibnamefont {Hudson}}, \bibinfo {author} {\bibfnamefont
  {H.}~\bibnamefont {Eisaki}}, \bibinfo {author} {\bibfnamefont
  {S.}~\bibnamefont {Uchida}}, \ and\ \bibinfo {author} {\bibfnamefont {J.~C.}\
  \bibnamefont {Davis}},\ }\bibfield  {title} {\enquote {\bibinfo {title}
  {Coincidence of checkerboard charge order and antinodal state decoherence in
  strongly underdoped superconducting {Bi$_2$Sr$_2$CaCu$_2$O$_{8+\delta}$}},}\
  }\href {\doibase 10.1103/PhysRevLett.94.197005} {\bibfield  {journal}
  {\bibinfo  {journal} {Phys. Rev. Lett.}\ }\textbf {\bibinfo {volume} {94}},\
  \bibinfo {pages} {197005} (\bibinfo {year} {2005}{\natexlab{a}})}\BibitemShut
  {NoStop}%
\bibitem [{\citenamefont {Fei}\ \emph {et~al.}(2018)\citenamefont {Fei},
  \citenamefont {Bu}, \citenamefont {Zhang}, \citenamefont {Zheng},
  \citenamefont {Sun}, \citenamefont {Ding}, \citenamefont {Zhou},\ and\
  \citenamefont {Yin}}]{FeiArXiv2018}%
  \BibitemOpen
  \bibfield  {author} {\bibinfo {author} {\bibfnamefont {Y.}~\bibnamefont
  {Fei}}, \bibinfo {author} {\bibfnamefont {K.}~\bibnamefont {Bu}}, \bibinfo
  {author} {\bibfnamefont {W.}~\bibnamefont {Zhang}}, \bibinfo {author}
  {\bibfnamefont {Y.}~\bibnamefont {Zheng}}, \bibinfo {author} {\bibfnamefont
  {X.}~\bibnamefont {Sun}}, \bibinfo {author} {\bibfnamefont {Y.}~\bibnamefont
  {Ding}}, \bibinfo {author} {\bibfnamefont {X.}~\bibnamefont {Zhou}}, \ and\
  \bibinfo {author} {\bibfnamefont {Y.}~\bibnamefont {Yin}},\ }\bibfield
  {title} {\enquote {\bibinfo {title} {Electronic effect of doped oxygen atoms
  in {Bi2201} superconductors determined by scanning tunneling microscopy},}\
  }\href {\doibase 10.1007/s11433-018-9276-5} {\bibfield  {journal} {\bibinfo
  {journal} {Sci. China Phys. Mech. Astron.}\ }\textbf {\bibinfo {volume}
  {61}},\ \bibinfo {pages} {127404} (\bibinfo {year} {2018})}\BibitemShut
  {NoStop}%
\bibitem [{\citenamefont {McElroy}\ \emph
  {et~al.}(2005{\natexlab{b}})\citenamefont {McElroy}, \citenamefont {Lee},
  \citenamefont {Slezak}, \citenamefont {Lee}, \citenamefont {Eisaki},
  \citenamefont {Uchida},\ and\ \citenamefont {Davis}}]{McElroyScience2005}%
  \BibitemOpen
  \bibfield  {author} {\bibinfo {author} {\bibfnamefont {K.}~\bibnamefont
  {McElroy}}, \bibinfo {author} {\bibfnamefont {J.}~\bibnamefont {Lee}},
  \bibinfo {author} {\bibfnamefont {J.~A.}\ \bibnamefont {Slezak}}, \bibinfo
  {author} {\bibfnamefont {D.-H.}\ \bibnamefont {Lee}}, \bibinfo {author}
  {\bibfnamefont {H.}~\bibnamefont {Eisaki}}, \bibinfo {author} {\bibfnamefont
  {S.}~\bibnamefont {Uchida}}, \ and\ \bibinfo {author} {\bibfnamefont {J.~C.}\
  \bibnamefont {Davis}},\ }\bibfield  {title} {\enquote {\bibinfo {title}
  {Atomic-scale sources and mechanism of nanoscale electronic disorder in
  {Bi$_2$Sr$_2$CaCu$_2$O$_{8+\delta}$}},}\ }\href {\doibase
  10.1126/science.1113095} {\bibfield  {journal} {\bibinfo  {journal}
  {Science}\ }\textbf {\bibinfo {volume} {309}},\ \bibinfo {pages} {1048--1052}
  (\bibinfo {year} {2005}{\natexlab{b}})}\BibitemShut {NoStop}%
\bibitem [{\citenamefont {Piriou}\ \emph {et~al.}(2011)\citenamefont {Piriou},
  \citenamefont {Jenkins}, \citenamefont {Berthod}, \citenamefont
  {Maggio-Aprile},\ and\ \citenamefont {Fischer}}]{PiriouNatCom2011}%
  \BibitemOpen
  \bibfield  {author} {\bibinfo {author} {\bibfnamefont {A.}~\bibnamefont
  {Piriou}}, \bibinfo {author} {\bibfnamefont {N.}~\bibnamefont {Jenkins}},
  \bibinfo {author} {\bibfnamefont {C.}~\bibnamefont {Berthod}}, \bibinfo
  {author} {\bibfnamefont {I.}~\bibnamefont {Maggio-Aprile}}, \ and\ \bibinfo
  {author} {\bibfnamefont {{\O}.}~\bibnamefont {Fischer}},\ }\bibfield  {title}
  {\enquote {\bibinfo {title} {First direct observation of the {Van Hove}
  singularity in the tunnelling spectra of cuprates},}\ }\href {\doibase
  10.1038/ncomms1229} {\bibfield  {journal} {\bibinfo  {journal} {Nat.
  Commun.}\ }\textbf {\bibinfo {volume} {2}},\ \bibinfo {pages} {221} (\bibinfo
  {year} {2011})}\BibitemShut {NoStop}%
\bibitem [{\citenamefont {Kinoda}\ \emph {et~al.}(2005)\citenamefont {Kinoda},
  \citenamefont {Mashima}, \citenamefont {Shimizu}, \citenamefont {Shimoyama},
  \citenamefont {Kishio},\ and\ \citenamefont {Hasegawa}}]{KinodaPRB2005}%
  \BibitemOpen
  \bibfield  {author} {\bibinfo {author} {\bibfnamefont {G.}~\bibnamefont
  {Kinoda}}, \bibinfo {author} {\bibfnamefont {H.}~\bibnamefont {Mashima}},
  \bibinfo {author} {\bibfnamefont {K.}~\bibnamefont {Shimizu}}, \bibinfo
  {author} {\bibfnamefont {J.}~\bibnamefont {Shimoyama}}, \bibinfo {author}
  {\bibfnamefont {K.}~\bibnamefont {Kishio}}, \ and\ \bibinfo {author}
  {\bibfnamefont {T.}~\bibnamefont {Hasegawa}},\ }\bibfield  {title} {\enquote
  {\bibinfo {title} {Direct determination of localized impurity levels located
  in the blocking layers of {Bi$_2$Sr$_2$CaCu$_2$O$_y$} using scanning
  tunneling microscopy/spectroscopy},}\ }\href {\doibase
  10.1103/PhysRevB.71.020502} {\bibfield  {journal} {\bibinfo  {journal} {Phys.
  Rev. B}\ }\textbf {\bibinfo {volume} {71}},\ \bibinfo {pages} {020502(R)}
  (\bibinfo {year} {2005})}\BibitemShut {NoStop}%
\bibitem [{\citenamefont {Miyakawa}\ \emph {et~al.}(1998)\citenamefont
  {Miyakawa}, \citenamefont {Guptasarma}, \citenamefont {Zasadzinski},
  \citenamefont {Hinks},\ and\ \citenamefont {Gray}}]{MiyakawaPRL1998}%
  \BibitemOpen
  \bibfield  {author} {\bibinfo {author} {\bibfnamefont {N.}~\bibnamefont
  {Miyakawa}}, \bibinfo {author} {\bibfnamefont {P.}~\bibnamefont
  {Guptasarma}}, \bibinfo {author} {\bibfnamefont {J.~F.}\ \bibnamefont
  {Zasadzinski}}, \bibinfo {author} {\bibfnamefont {D.~G.}\ \bibnamefont
  {Hinks}}, \ and\ \bibinfo {author} {\bibfnamefont {K.~E.}\ \bibnamefont
  {Gray}},\ }\bibfield  {title} {\enquote {\bibinfo {title} {Strong dependence
  of the superconducting gap on oxygen doping from tunneling measurements on
  {Bi$_2$Sr$_2$CaCu$_2$O$_{8-\delta}$}},}\ }\href {\doibase
  10.1103/PhysRevLett.80.157} {\bibfield  {journal} {\bibinfo  {journal} {Phys.
  Rev. Lett.}\ }\textbf {\bibinfo {volume} {80}},\ \bibinfo {pages} {157--160}
  (\bibinfo {year} {1998})}\BibitemShut {NoStop}%
\bibitem [{\citenamefont {White}\ \emph {et~al.}(1996)\citenamefont {White},
  \citenamefont {Shen}, \citenamefont {Kim}, \citenamefont {Harris},
  \citenamefont {Loeser}, \citenamefont {Fournier},\ and\ \citenamefont
  {Kapitulnik}}]{WhitePRB1996}%
  \BibitemOpen
  \bibfield  {author} {\bibinfo {author} {\bibfnamefont {P.~J.}\ \bibnamefont
  {White}}, \bibinfo {author} {\bibfnamefont {Z.-X.}\ \bibnamefont {Shen}},
  \bibinfo {author} {\bibfnamefont {C.}~\bibnamefont {Kim}}, \bibinfo {author}
  {\bibfnamefont {J.~M.}\ \bibnamefont {Harris}}, \bibinfo {author}
  {\bibfnamefont {A.~G.}\ \bibnamefont {Loeser}}, \bibinfo {author}
  {\bibfnamefont {P.}~\bibnamefont {Fournier}}, \ and\ \bibinfo {author}
  {\bibfnamefont {A.}~\bibnamefont {Kapitulnik}},\ }\bibfield  {title}
  {\enquote {\bibinfo {title} {Rapid suppression of the superconducting gap in
  overdoped {Bi$_2$Sr$_2$CaCu$_2$O$_{8+\delta}$}},}\ }\href {\doibase
  10.1103/PhysRevB.54.R15669} {\bibfield  {journal} {\bibinfo  {journal} {Phys.
  Rev. B}\ }\textbf {\bibinfo {volume} {54}},\ \bibinfo {pages} {R15669(R)}
  (\bibinfo {year} {1996})}\BibitemShut {NoStop}%
\bibitem [{\citenamefont {Harris}\ \emph {et~al.}(1996)\citenamefont {Harris},
  \citenamefont {Shen}, \citenamefont {White}, \citenamefont {Marshall},
  \citenamefont {Schabel}, \citenamefont {Eckstein},\ and\ \citenamefont
  {Bozovic}}]{HarrisPRB1996}%
  \BibitemOpen
  \bibfield  {author} {\bibinfo {author} {\bibfnamefont {J.~M.}\ \bibnamefont
  {Harris}}, \bibinfo {author} {\bibfnamefont {Z.-X.}\ \bibnamefont {Shen}},
  \bibinfo {author} {\bibfnamefont {P.~J.}\ \bibnamefont {White}}, \bibinfo
  {author} {\bibfnamefont {D.~S.}\ \bibnamefont {Marshall}}, \bibinfo {author}
  {\bibfnamefont {M.~C.}\ \bibnamefont {Schabel}}, \bibinfo {author}
  {\bibfnamefont {J.~N.}\ \bibnamefont {Eckstein}}, \ and\ \bibinfo {author}
  {\bibfnamefont {I.}~\bibnamefont {Bozovic}},\ }\bibfield  {title} {\enquote
  {\bibinfo {title} {Anomalous superconducting state gap size versus {$T_c$}
  behavior in underdoped {Bi$_2$Sr$_2$Ca$_{1-x}$Dy$_x$Cu$_2$O$_{8+\delta}$}},}\
  }\href {\doibase 10.1103/PhysRevB.54.R15665} {\bibfield  {journal} {\bibinfo
  {journal} {Phys. Rev. B}\ }\textbf {\bibinfo {volume} {54}},\ \bibinfo
  {pages} {R15665(R)} (\bibinfo {year} {1996})}\BibitemShut {NoStop}%
\bibitem [{\citenamefont {Ding}\ \emph {et~al.}(2001)\citenamefont {Ding},
  \citenamefont {Engelbrecht}, \citenamefont {Wang}, \citenamefont {Campuzano},
  \citenamefont {Wang}, \citenamefont {Yang}, \citenamefont {Rogan},
  \citenamefont {Takahashi}, \citenamefont {Kadowaki},\ and\ \citenamefont
  {Hinks}}]{DingPRL2001}%
  \BibitemOpen
  \bibfield  {author} {\bibinfo {author} {\bibfnamefont {H.}~\bibnamefont
  {Ding}}, \bibinfo {author} {\bibfnamefont {J.~R.}\ \bibnamefont
  {Engelbrecht}}, \bibinfo {author} {\bibfnamefont {Z.}~\bibnamefont {Wang}},
  \bibinfo {author} {\bibfnamefont {J.~C.}\ \bibnamefont {Campuzano}}, \bibinfo
  {author} {\bibfnamefont {S.-C.}\ \bibnamefont {Wang}}, \bibinfo {author}
  {\bibfnamefont {H.-B.}\ \bibnamefont {Yang}}, \bibinfo {author}
  {\bibfnamefont {R.}~\bibnamefont {Rogan}}, \bibinfo {author} {\bibfnamefont
  {T.}~\bibnamefont {Takahashi}}, \bibinfo {author} {\bibfnamefont
  {K.}~\bibnamefont {Kadowaki}}, \ and\ \bibinfo {author} {\bibfnamefont
  {D.~G.}\ \bibnamefont {Hinks}},\ }\bibfield  {title} {\enquote {\bibinfo
  {title} {Coherent quasiparticle weight and its connection to high-{$T_c$}
  superconductivity from angle-resolved photoemission},}\ }\href {\doibase
  10.1103/PhysRevLett.87.227001} {\bibfield  {journal} {\bibinfo  {journal}
  {Phys. Rev. Lett.}\ }\textbf {\bibinfo {volume} {87}},\ \bibinfo {pages}
  {227001} (\bibinfo {year} {2001})}\BibitemShut {NoStop}%
\bibitem [{\citenamefont {H\"{u}fner}\ \emph {et~al.}(2008)\citenamefont
  {H\"{u}fner}, \citenamefont {Hossain}, \citenamefont {Damascelli},\ and\
  \citenamefont {Sawatzky}}]{HufnerRPP2008}%
  \BibitemOpen
  \bibfield  {author} {\bibinfo {author} {\bibfnamefont {S.}~\bibnamefont
  {H\"{u}fner}}, \bibinfo {author} {\bibfnamefont {M.~A.}\ \bibnamefont
  {Hossain}}, \bibinfo {author} {\bibfnamefont {A.}~\bibnamefont {Damascelli}},
  \ and\ \bibinfo {author} {\bibfnamefont {G.~A.}\ \bibnamefont {Sawatzky}},\
  }\bibfield  {title} {\enquote {\bibinfo {title} {Two gaps make a
  high-temperature superconductor?}}\ }\href {\doibase
  10.1088/0034-4885/71/6/062501} {\bibfield  {journal} {\bibinfo  {journal}
  {Rep. Prog. Phys.}\ }\textbf {\bibinfo {volume} {71}},\ \bibinfo {pages}
  {062501} (\bibinfo {year} {2008})}\BibitemShut {NoStop}%
\bibitem [{\citenamefont {Fujita}\ \emph {et~al.}(2008)\citenamefont {Fujita},
  \citenamefont {Grigorenko}, \citenamefont {Lee}, \citenamefont {Wang},
  \citenamefont {Davis}, \citenamefont {Eisaki}, \citenamefont {Uchida},\ and\
  \citenamefont {Balatsky}}]{FujitaJPCS2008}%
  \BibitemOpen
  \bibfield  {author} {\bibinfo {author} {\bibfnamefont {K.}~\bibnamefont
  {Fujita}}, \bibinfo {author} {\bibfnamefont {I.}~\bibnamefont {Grigorenko}},
  \bibinfo {author} {\bibfnamefont {J.}~\bibnamefont {Lee}}, \bibinfo {author}
  {\bibfnamefont {M.}~\bibnamefont {Wang}}, \bibinfo {author} {\bibfnamefont
  {J.~C.}\ \bibnamefont {Davis}}, \bibinfo {author} {\bibfnamefont
  {H.}~\bibnamefont {Eisaki}}, \bibinfo {author} {\bibfnamefont
  {S.}~\bibnamefont {Uchida}}, \ and\ \bibinfo {author} {\bibfnamefont {A.~V.}\
  \bibnamefont {Balatsky}},\ }\bibfield  {title} {\enquote {\bibinfo {title}
  {{Bogoliubov} angle and visualization of particle--hole mixture in
  superconductors},}\ }\href {\doibase 10.1103/PhysRevB.78.054510} {\bibfield
  {journal} {\bibinfo  {journal} {Phys. Rev. B}\ }\textbf {\bibinfo {volume}
  {78}},\ \bibinfo {pages} {054510} (\bibinfo {year} {2008})}\BibitemShut
  {NoStop}%
\bibitem [{\citenamefont {Wang}\ and\ \citenamefont {Lee}(2003)}]{WangPRB2003}%
  \BibitemOpen
  \bibfield  {author} {\bibinfo {author} {\bibfnamefont {Q.-H.}\ \bibnamefont
  {Wang}}\ and\ \bibinfo {author} {\bibfnamefont {D.-H.}\ \bibnamefont {Lee}},\
  }\bibfield  {title} {\enquote {\bibinfo {title} {Quasiparticle scattering
  interference in high-temperature superconductors},}\ }\href {\doibase
  10.1103/PhysRevB.67.020511} {\bibfield  {journal} {\bibinfo  {journal} {Phys.
  Rev. B}\ }\textbf {\bibinfo {volume} {67}},\ \bibinfo {pages} {020511(R)}
  (\bibinfo {year} {2003})}\BibitemShut {NoStop}%
\bibitem [{\citenamefont {Hoffman}\ \emph {et~al.}(2002)\citenamefont
  {Hoffman}, \citenamefont {McElroy}, \citenamefont {Lee}, \citenamefont
  {Lang}, \citenamefont {Eisaki}, \citenamefont {Uchida},\ and\ \citenamefont
  {Davis}}]{HoffmanScience2002}%
  \BibitemOpen
  \bibfield  {author} {\bibinfo {author} {\bibfnamefont {J.~E.}\ \bibnamefont
  {Hoffman}}, \bibinfo {author} {\bibfnamefont {K.}~\bibnamefont {McElroy}},
  \bibinfo {author} {\bibfnamefont {D.-H.}\ \bibnamefont {Lee}}, \bibinfo
  {author} {\bibfnamefont {K.~M.}\ \bibnamefont {Lang}}, \bibinfo {author}
  {\bibfnamefont {H.}~\bibnamefont {Eisaki}}, \bibinfo {author} {\bibfnamefont
  {S.}~\bibnamefont {Uchida}}, \ and\ \bibinfo {author} {\bibfnamefont {J.~C.}\
  \bibnamefont {Davis}},\ }\bibfield  {title} {\enquote {\bibinfo {title}
  {Imaging quasiparticle interference in
  {Bi$_2$Sr$_2$CaCu$_2$O$_{8+\delta}$}},}\ }\href {\doibase
  10.1126/science.1072640} {\bibfield  {journal} {\bibinfo  {journal}
  {Science}\ }\textbf {\bibinfo {volume} {297}},\ \bibinfo {pages} {1148--1151}
  (\bibinfo {year} {2002})}\BibitemShut {NoStop}%
\bibitem [{\citenamefont {McElroy}\ \emph {et~al.}(2003)\citenamefont
  {McElroy}, \citenamefont {Simmonds}, \citenamefont {Hoffman}, \citenamefont
  {Lee}, \citenamefont {Orenstein}, \citenamefont {Eisaki}, \citenamefont
  {Uchida},\ and\ \citenamefont {Davis}}]{McElroyNature2003}%
  \BibitemOpen
  \bibfield  {author} {\bibinfo {author} {\bibfnamefont {K.}~\bibnamefont
  {McElroy}}, \bibinfo {author} {\bibfnamefont {R.~W.}\ \bibnamefont
  {Simmonds}}, \bibinfo {author} {\bibfnamefont {J.~E.}\ \bibnamefont
  {Hoffman}}, \bibinfo {author} {\bibfnamefont {D.-H.}\ \bibnamefont {Lee}},
  \bibinfo {author} {\bibfnamefont {J.}~\bibnamefont {Orenstein}}, \bibinfo
  {author} {\bibfnamefont {H.}~\bibnamefont {Eisaki}}, \bibinfo {author}
  {\bibfnamefont {S.}~\bibnamefont {Uchida}}, \ and\ \bibinfo {author}
  {\bibfnamefont {J.~C.}\ \bibnamefont {Davis}},\ }\bibfield  {title} {\enquote
  {\bibinfo {title} {Relating atomic-scale electronic phenomena to wave-like
  quasiparticle states in superconducting
  {Bi$_2$Sr$_2$CaCu$_2$O$_{8+\delta}$}},}\ }\href {\doibase
  10.1038/nature01496} {\bibfield  {journal} {\bibinfo  {journal} {Nature}\
  }\textbf {\bibinfo {volume} {422}},\ \bibinfo {pages} {592--596} (\bibinfo
  {year} {2003})}\BibitemShut {NoStop}%
\bibitem [{\citenamefont {Hanaguri}\ \emph {et~al.}(2007)\citenamefont
  {Hanaguri}, \citenamefont {Kohsaka}, \citenamefont {Davis}, \citenamefont
  {Lupien}, \citenamefont {Yamada}, \citenamefont {Azuma}, \citenamefont
  {Takano}, \citenamefont {Ohishi}, \citenamefont {Ono},\ and\ \citenamefont
  {Takagi}}]{HanaguriNatPhys2007}%
  \BibitemOpen
  \bibfield  {author} {\bibinfo {author} {\bibfnamefont {T.}~\bibnamefont
  {Hanaguri}}, \bibinfo {author} {\bibfnamefont {Y.}~\bibnamefont {Kohsaka}},
  \bibinfo {author} {\bibfnamefont {J.~C.}\ \bibnamefont {Davis}}, \bibinfo
  {author} {\bibfnamefont {C.}~\bibnamefont {Lupien}}, \bibinfo {author}
  {\bibfnamefont {I.}~\bibnamefont {Yamada}}, \bibinfo {author} {\bibfnamefont
  {M.}~\bibnamefont {Azuma}}, \bibinfo {author} {\bibfnamefont
  {M.}~\bibnamefont {Takano}}, \bibinfo {author} {\bibfnamefont
  {K.}~\bibnamefont {Ohishi}}, \bibinfo {author} {\bibfnamefont
  {M.}~\bibnamefont {Ono}}, \ and\ \bibinfo {author} {\bibfnamefont
  {H.}~\bibnamefont {Takagi}},\ }\bibfield  {title} {\enquote {\bibinfo {title}
  {Quasiparticle interference and superconducting gap in
  {Ca$_{2−x}$Na$_x$CuO$_2$Cl$_2$}},}\ }\href {\doibase 10.1038/nphys753}
  {\bibfield  {journal} {\bibinfo  {journal} {Nat. Phys.}\ }\textbf {\bibinfo
  {volume} {3}},\ \bibinfo {pages} {865--871} (\bibinfo {year}
  {2007})}\BibitemShut {NoStop}%
\bibitem [{\citenamefont {Fujita}\ \emph
  {et~al.}(2014{\natexlab{b}})\citenamefont {Fujita}, \citenamefont {Hamidian},
  \citenamefont {Edkins}, \citenamefont {Kim}, \citenamefont {Kohsaka},
  \citenamefont {Azuma}, \citenamefont {Takano}, \citenamefont {Takagi},
  \citenamefont {Eisaki}, \citenamefont {Uchida}, \citenamefont {Allais},
  \citenamefont {Lawler}, \citenamefont {Kim}, \citenamefont {Sachdev},\ and\
  \citenamefont {Davis}}]{FujitaPNAS2014}%
  \BibitemOpen
  \bibfield  {author} {\bibinfo {author} {\bibfnamefont {K.}~\bibnamefont
  {Fujita}}, \bibinfo {author} {\bibfnamefont {M.~H.}\ \bibnamefont
  {Hamidian}}, \bibinfo {author} {\bibfnamefont {S.~D.}\ \bibnamefont
  {Edkins}}, \bibinfo {author} {\bibfnamefont {C.~K.}\ \bibnamefont {Kim}},
  \bibinfo {author} {\bibfnamefont {Y.}~\bibnamefont {Kohsaka}}, \bibinfo
  {author} {\bibfnamefont {M.}~\bibnamefont {Azuma}}, \bibinfo {author}
  {\bibfnamefont {M.}~\bibnamefont {Takano}}, \bibinfo {author} {\bibfnamefont
  {H.}~\bibnamefont {Takagi}}, \bibinfo {author} {\bibfnamefont
  {H.}~\bibnamefont {Eisaki}}, \bibinfo {author} {\bibfnamefont {S.-I.}\
  \bibnamefont {Uchida}}, \bibinfo {author} {\bibfnamefont {A.}~\bibnamefont
  {Allais}}, \bibinfo {author} {\bibfnamefont {M.~J.}\ \bibnamefont {Lawler}},
  \bibinfo {author} {\bibfnamefont {E.-A.}\ \bibnamefont {Kim}}, \bibinfo
  {author} {\bibfnamefont {S.}~\bibnamefont {Sachdev}}, \ and\ \bibinfo
  {author} {\bibfnamefont {J.~C.~S.}\ \bibnamefont {Davis}},\ }\bibfield
  {title} {\enquote {\bibinfo {title} {Direct phase-sensitive identification of
  a $d$-form factor density wave in underdoped cuprates},}\ }\href {\doibase
  10.1073/pnas.1406297111} {\bibfield  {journal} {\bibinfo  {journal} {Proc.
  Natl. Acad. Sci. USA}\ }\textbf {\bibinfo {volume} {111}},\ \bibinfo {pages}
  {E3026--E3032} (\bibinfo {year} {2014}{\natexlab{b}})}\BibitemShut {NoStop}%
\bibitem [{\citenamefont {Hamidian}\ \emph {et~al.}(2016)\citenamefont
  {Hamidian}, \citenamefont {Edkins}, \citenamefont {Kim}, \citenamefont
  {Davis}, \citenamefont {Mackenzie}, \citenamefont {Eisaki}, \citenamefont
  {Uchida}, \citenamefont {Lawler}, \citenamefont {Kim}, \citenamefont
  {Sachdev},\ and\ \citenamefont {Fujita}}]{HamidianNatPhys2016}%
  \BibitemOpen
  \bibfield  {author} {\bibinfo {author} {\bibfnamefont {M.~H.}\ \bibnamefont
  {Hamidian}}, \bibinfo {author} {\bibfnamefont {S.~D.}\ \bibnamefont
  {Edkins}}, \bibinfo {author} {\bibfnamefont {C.~K.}\ \bibnamefont {Kim}},
  \bibinfo {author} {\bibfnamefont {J.~C.}\ \bibnamefont {Davis}}, \bibinfo
  {author} {\bibfnamefont {A.~P.}\ \bibnamefont {Mackenzie}}, \bibinfo {author}
  {\bibfnamefont {H.}~\bibnamefont {Eisaki}}, \bibinfo {author} {\bibfnamefont
  {S.}~\bibnamefont {Uchida}}, \bibinfo {author} {\bibfnamefont {M.~J.}\
  \bibnamefont {Lawler}}, \bibinfo {author} {\bibfnamefont {E.-A.}\
  \bibnamefont {Kim}}, \bibinfo {author} {\bibfnamefont {S.}~\bibnamefont
  {Sachdev}}, \ and\ \bibinfo {author} {\bibfnamefont {K.}~\bibnamefont
  {Fujita}},\ }\bibfield  {title} {\enquote {\bibinfo {title} {Atomic-scale
  electronic structure of the cuprate $d$-symmetry form factor density wave
  state},}\ }\href {\doibase 10.1038/nphys3519} {\bibfield  {journal} {\bibinfo
   {journal} {Nat. Phys.}\ }\textbf {\bibinfo {volume} {12}},\ \bibinfo {pages}
  {150--156} (\bibinfo {year} {2016})}\BibitemShut {NoStop}%
\bibitem [{\citenamefont {Mesaros}\ \emph {et~al.}(2016)\citenamefont
  {Mesaros}, \citenamefont {Fujita}, \citenamefont {Edkins}, \citenamefont
  {Hamidian}, \citenamefont {Eisaki}, \citenamefont {Uchida}, \citenamefont
  {Davis}, \citenamefont {Lawler},\ and\ \citenamefont
  {Kim}}]{MesarosPNAS2016}%
  \BibitemOpen
  \bibfield  {author} {\bibinfo {author} {\bibfnamefont {A.}~\bibnamefont
  {Mesaros}}, \bibinfo {author} {\bibfnamefont {K.}~\bibnamefont {Fujita}},
  \bibinfo {author} {\bibfnamefont {S.~D.}\ \bibnamefont {Edkins}}, \bibinfo
  {author} {\bibfnamefont {M.~H.}\ \bibnamefont {Hamidian}}, \bibinfo {author}
  {\bibfnamefont {H.}~\bibnamefont {Eisaki}}, \bibinfo {author} {\bibfnamefont
  {S.-I.}\ \bibnamefont {Uchida}}, \bibinfo {author} {\bibfnamefont {J.~C.~S.}\
  \bibnamefont {Davis}}, \bibinfo {author} {\bibfnamefont {M.~J.}\ \bibnamefont
  {Lawler}}, \ and\ \bibinfo {author} {\bibfnamefont {E.-A.}\ \bibnamefont
  {Kim}},\ }\bibfield  {title} {\enquote {\bibinfo {title} {Commensurate
  4$a_0$-period charge density modulations throughout the
  {Bi$_2$Sr$_2$CaCu$_2$O$_{8+x}$} pseudogap regime},}\ }\href {\doibase
  10.1073/pnas.1614247113} {\bibfield  {journal} {\bibinfo  {journal} {Proc.
  Natl. Acad. Sci. USA}\ }\textbf {\bibinfo {volume} {113}},\ \bibinfo {pages}
  {12661--12666} (\bibinfo {year} {2016})}\BibitemShut {NoStop}%
\bibitem [{\citenamefont {Zhang}\ \emph {et~al.}(2018)\citenamefont {Zhang},
  \citenamefont {Mesaros}, \citenamefont {Fujita}, \citenamefont {Edkins},
  \citenamefont {Hamidian}, \citenamefont {Ch'ng}, \citenamefont {Eisaki},
  \citenamefont {Uchida}, \citenamefont {Davis}, \citenamefont {Khatami},\ and\
  \citenamefont {Kim}}]{ZhangArXiv2018}%
  \BibitemOpen
  \bibfield  {author} {\bibinfo {author} {\bibfnamefont {Y.}~\bibnamefont
  {Zhang}}, \bibinfo {author} {\bibfnamefont {A.}~\bibnamefont {Mesaros}},
  \bibinfo {author} {\bibfnamefont {K.}~\bibnamefont {Fujita}}, \bibinfo
  {author} {\bibfnamefont {S.~D.}\ \bibnamefont {Edkins}}, \bibinfo {author}
  {\bibfnamefont {M.~H.}\ \bibnamefont {Hamidian}}, \bibinfo {author}
  {\bibfnamefont {K.}~\bibnamefont {Ch'ng}}, \bibinfo {author} {\bibfnamefont
  {H.}~\bibnamefont {Eisaki}}, \bibinfo {author} {\bibfnamefont
  {S.}~\bibnamefont {Uchida}}, \bibinfo {author} {\bibfnamefont {J.~C.~S.}\
  \bibnamefont {Davis}}, \bibinfo {author} {\bibfnamefont {E.}~\bibnamefont
  {Khatami}}, \ and\ \bibinfo {author} {\bibfnamefont {E.-A.}\ \bibnamefont
  {Kim}},\ }\bibfield  {title} {\enquote {\bibinfo {title} {Machine learning in
  electronic quantum matter imaging experiments},}\ }\href@noop {} {\
  (\bibinfo {year} {2018})},\ \Eprint {http://arxiv.org/abs/1808.00479}
  {arXiv:1808.00479} \BibitemShut {NoStop}%
\bibitem [{\citenamefont {Zhao}\ \emph {et~al.}(2019)\citenamefont {Zhao},
  \citenamefont {Ren}, \citenamefont {Rachmilowitz}, \citenamefont
  {Schneeloch}, \citenamefont {Zhong}, \citenamefont {Gu}, \citenamefont
  {Wang},\ and\ \citenamefont {Zeljkovic}}]{ZhaoNatMat2018}%
  \BibitemOpen
  \bibfield  {author} {\bibinfo {author} {\bibfnamefont {H.}~\bibnamefont
  {Zhao}}, \bibinfo {author} {\bibfnamefont {Z.}~\bibnamefont {Ren}}, \bibinfo
  {author} {\bibfnamefont {B.}~\bibnamefont {Rachmilowitz}}, \bibinfo {author}
  {\bibfnamefont {J.}~\bibnamefont {Schneeloch}}, \bibinfo {author}
  {\bibfnamefont {R.}~\bibnamefont {Zhong}}, \bibinfo {author} {\bibfnamefont
  {G.}~\bibnamefont {Gu}}, \bibinfo {author} {\bibfnamefont {Z.}~\bibnamefont
  {Wang}}, \ and\ \bibinfo {author} {\bibfnamefont {I.}~\bibnamefont
  {Zeljkovic}},\ }\bibfield  {title} {\enquote {\bibinfo {title} {Charge-stripe
  crystal phase in an insulating cuprate},}\ }\href {\doibase
  10.1038/s41563-018-0243-x} {\bibfield  {journal} {\bibinfo  {journal} {Nat.
  Mater.}\ }\textbf {\bibinfo {volume} {18}},\ \bibinfo {pages} {103--107}
  (\bibinfo {year} {2019})}\BibitemShut {NoStop}%
\bibitem [{Note3()}]{Note3}%
  \BibitemOpen
  \bibinfo {note} {Ref.~\cite {CominScience2014} concludes that $\protect
  \mathaccentV {bar}016{Q}_{\protect \mathrm {DW}}$ is explained by a FS
  instability, apparently in contradiction to the findings reported herein.
  However, Ref.~\cite {CominScience2014} models an incommensurate $\protect
  \mathaccentV {bar}016{Q}_{\protect \mathrm {DW}}$ in the presence of Fermi
  arcs, where the renormalization associated with the arc phenomenology
  generates a $Q_{\protect \mathrm {HS}}$ significantly larger than
  $Q_{\protect \mathrm {AFZB}}$}\BibitemShut {NoStop}%
\bibitem [{\citenamefont {Chowdhury}\ and\ \citenamefont
  {Sachdev}(2014)}]{ChowdhuryPRB2014}%
  \BibitemOpen
  \bibfield  {author} {\bibinfo {author} {\bibfnamefont {D.}~\bibnamefont
  {Chowdhury}}\ and\ \bibinfo {author} {\bibfnamefont {S.}~\bibnamefont
  {Sachdev}},\ }\bibfield  {title} {\enquote {\bibinfo {title} {Density-wave
  instabilities of fractionalized {Fermi} liquids},}\ }\href {\doibase
  10.1103/PhysRevB.90.245136} {\bibfield  {journal} {\bibinfo  {journal} {Phys.
  Rev. B}\ }\textbf {\bibinfo {volume} {90}},\ \bibinfo {pages} {245136}
  (\bibinfo {year} {2014})}\BibitemShut {NoStop}%
\bibitem [{\citenamefont {Bak}(1982)}]{Bak1982RPP}%
  \BibitemOpen
  \bibfield  {author} {\bibinfo {author} {\bibfnamefont {P.}~\bibnamefont
  {Bak}},\ }\bibfield  {title} {\enquote {\bibinfo {title} {Commensurate
  phases, incommensurate phases and the devil's staircase},}\ }\href {\doibase
  10.1088/0034-4885/45/6/001} {\bibfield  {journal} {\bibinfo  {journal} {Rep.
  Prog. Phys.}\ }\textbf {\bibinfo {volume} {45}},\ \bibinfo {pages} {587}
  (\bibinfo {year} {1982})}\BibitemShut {NoStop}%
\bibitem [{\citenamefont {Lawler}\ \emph {et~al.}(2010)\citenamefont {Lawler},
  \citenamefont {Fujita}, \citenamefont {Lee}, \citenamefont {Schmidt},
  \citenamefont {Kohsaka}, \citenamefont {Kim}, \citenamefont {Eisaki},
  \citenamefont {Uchida}, \citenamefont {Davis}, \citenamefont {Sethna},\ and\
  \citenamefont {Kim}}]{LawlerNature2010}%
  \BibitemOpen
  \bibfield  {author} {\bibinfo {author} {\bibfnamefont {M.~J.}\ \bibnamefont
  {Lawler}}, \bibinfo {author} {\bibfnamefont {K.}~\bibnamefont {Fujita}},
  \bibinfo {author} {\bibfnamefont {J.}~\bibnamefont {Lee}}, \bibinfo {author}
  {\bibfnamefont {A.~R.}\ \bibnamefont {Schmidt}}, \bibinfo {author}
  {\bibfnamefont {Y.}~\bibnamefont {Kohsaka}}, \bibinfo {author} {\bibfnamefont
  {C.~K.}\ \bibnamefont {Kim}}, \bibinfo {author} {\bibfnamefont
  {H.}~\bibnamefont {Eisaki}}, \bibinfo {author} {\bibfnamefont
  {S.}~\bibnamefont {Uchida}}, \bibinfo {author} {\bibfnamefont {J.~C.}\
  \bibnamefont {Davis}}, \bibinfo {author} {\bibfnamefont {J.~P.}\ \bibnamefont
  {Sethna}}, \ and\ \bibinfo {author} {\bibfnamefont {E.-A.}\ \bibnamefont
  {Kim}},\ }\bibfield  {title} {\enquote {\bibinfo {title} {Intra-unit-cell
  electronic nematicity of the high-{$T_c$} copper-oxide pseudogap states},}\
  }\href {\doibase 10.1038/nature09169} {\bibfield  {journal} {\bibinfo
  {journal} {Nature}\ }\textbf {\bibinfo {volume} {466}},\ \bibinfo {pages}
  {347--351} (\bibinfo {year} {2010})}\BibitemShut {NoStop}%
\bibitem [{\citenamefont {Kondo}\ \emph {et~al.}(2009)\citenamefont {Kondo},
  \citenamefont {Khasanov}, \citenamefont {Takeuchi}, \citenamefont
  {Schmalian},\ and\ \citenamefont {Kaminski}}]{KondoNature2009}%
  \BibitemOpen
  \bibfield  {author} {\bibinfo {author} {\bibfnamefont {T.}~\bibnamefont
  {Kondo}}, \bibinfo {author} {\bibfnamefont {R.}~\bibnamefont {Khasanov}},
  \bibinfo {author} {\bibfnamefont {T.}~\bibnamefont {Takeuchi}}, \bibinfo
  {author} {\bibfnamefont {J.}~\bibnamefont {Schmalian}}, \ and\ \bibinfo
  {author} {\bibfnamefont {A.}~\bibnamefont {Kaminski}},\ }\bibfield  {title}
  {\enquote {\bibinfo {title} {Competition between the pseudogap and
  superconductivity in the high-{$T_c$} copper oxides},}\ }\href {\doibase
  10.1038/nature07644} {\bibfield  {journal} {\bibinfo  {journal} {Nature}\
  }\textbf {\bibinfo {volume} {457}},\ \bibinfo {pages} {296--300} (\bibinfo
  {year} {2009})}\BibitemShut {NoStop}%
\end{thebibliography}%

\end{document}